\documentclass[useAMS,usenatbib]{mn2e}
\usepackage{graphicx}
\usepackage{setspace}
\usepackage{natbib}
\usepackage{color}
\usepackage{amsmath}
\usepackage{times}
\usepackage{aas_macros}

\title[Mismatch and Misalignment]{Mismatch and Misalignment: Dark
  Haloes and Satellites of Disc Galaxies}

\author[A. J. Deason et al.]
{A.J. Deason$^{1}$\thanks{E-mail:ajd75@ast.cam.ac.uk},
  I.G. McCarthy$^{1,2,3}$, A.S. Font$^{1}$, N. W. Evans$^{1}$,  
  C.S Frenk$^4$, V. Belokurov$^{1}$,
\newauthor N.I. Libeskind$^5$,
  R.A. Crain$^6$, T. Theuns$^{4,7}$ \\
  $^{1}$Institute of Astronomy, Madingley Rd, Cambridge, CB3 0HA\\
  $^{2}$Kavli Institute for Cosmology, University of Cambridge,
  Madingley Road, Cambridge, CB3 OHA\\
  $^{3}$Astrophysics Group, Cavendish Laboratory, JJ Thomson Avenue,
  Cambridge, CB3 0HE\\
  $^{4}$Institute for Computational Cosmology, Department of Physics,
  University of Durham, South Road, Durham DH1 3LE\\
  $^{5}$ Astrophysikalisches Institut Potsdam, An der Sternwarte 16,
  D-14482 Potsdam, Germany\\
  $^{6}$Centre for Astrophysics \& Supercomputing, Swinburne
  University of Technology, Hawthorn, Victoria 3122, Australia\\
  $^{7}$Department of Physics, University of Antwerp, Campus
  Groenenborger, Groenenborgerlaan 171, B-2020 Antwerp, Belgium}

\begin{document}


\date{June 2011}

\pagerange{\pageref{firstpage}--\pageref{lastpage}} \pubyear{2010}

\maketitle

\label{firstpage}

\begin{abstract} 
 We study the phase-space distribution of satellite galaxies
  associated with late-type galaxies in the \textsc{gimic} suite of
  simulations. \textsc{gimic} consists of re-simulations of 5
  cosmologically representative regions from the \textit{Millennium
    simulation}, which have higher resolution and incorporate baryonic
  physics. Whilst the disc of the galaxy is well aligned with
  the inner regions ($r \sim 0.1r_{200}$) of the dark matter halo,
  both in shape and angular momentum, there can be substantial
  misalignments at larger radii ($r \sim r_{200}$). Misalignments of $
  > 45^\circ$ are seen in $\sim 30\%$ of our sample. We find that the
  satellite population aligns with the shape (and angular momentum) of
  the outer dark matter halo. However, the alignment with the galaxy is weak owing to the mismatch between the disc and dark
  matter halo. Roughly $20\%$ of the satellite systems with ten
  bright galaxies within $r_{200}$ exhibit a polar spatial alignment
  with respect to the galaxy --- an orientation reminiscent of
  the classical satellites of the Milky Way. We find that a small fraction ($\sim
  10\%$) of satellite systems show evidence for rotational support
  which we attribute to group infall. There is a bias towards satellites on prograde
  orbits relative to the spin of the dark matter halo (and to a lesser
  extent with the angular momentum of the disc). This preference
  towards co-rotation is stronger in the inner regions of the
  halo where the most massive satellites accreted at
  relatively early times are located. 

  We attribute the
  anisotropic spatial distribution and angular momentum bias of the
  satellites at $z=0$ to their directional accretion along the major
  axes of the dark matter halo. The satellite
  galaxies have been accreted relatively recently compared to the dark
  matter mass and have experienced less phase-mixing and relaxation
  --- the memory of their accretion history can remain intact to
  $z=0$. Understanding the phase-space distribution of the $z=0$
  satellite population is key for studies that estimate the host halo
  mass from the line of sight velocities and projected positions of
  satellite galaxies. We quantify the effects of such systematics in
  estimates of the host halo mass from the satellite population.
\end{abstract}


\section{Introduction}

Studies of local galaxies, such as the Milky Way and M31, can
potentially provide us with the missing link between cosmological
structure formation and the complex baryonic processes that help
shape the galaxies we observe today. Any acceptable cosmological model
must, in addition to satisfying the requirements of large scale
structure, account for the small scale detail exhibited by our own
Milky Way galaxy and others.

For some time it has been known that the eleven classical satellites
of the Milky Way define a highly inclined plane relative to the disc
of the Galaxy (\citealt{lyndenbell76}). Early work by
\cite{holmberg69} and \cite{zaritsky97} found similar alignments in
external galaxies whereby the satellites tend to avoid the equatorial
regions of the parent light distribution. However, more recent work
using the 2 degree field Galaxy Redshift Survey and the Sloan Digital
Sky Survey (SDSS) find that the opposite trend is true; satellites
tend to avoid the polar regions of the light distribution
(\citealt{sales04}; \citealt{brainerd05};
\citealt{yang06}). \cite{agustsson10} find that the spatial
orientation of satellite galaxies depends on the type of host galaxy;
satellites of red galaxies are found preferentially near the major
axes of their hosts while satellites of blue galaxies show little or
no spatial ansiotropy. The anisotropic nature of the spatial
distribution of satellite galaxies has led some authors to postulate a
discrepancy between observational constraints and models adopting the
$\Lambda$CDM framework (e.g. \citealt{kroupa05}). However, numerical
studies have shown that preferential alignments are naturally produced
in the simulations (\citealt{kang05}; \citealt{libeskind05};
\citealt{zentner05}). The satellites tend to align with the major axis
of the dark matter halo, but an extrapolation to the relation with the
light distribution is not straightforward in models which do not
follow the evolution of baryonic matter.

The radial velocities and available proper motions of the classical
Milky Way satellites hint at the presence of coherent motion
(\citealt{lyndenbell95}; \citealt{metz08}). In fact, \cite{metz08}
find that at least 3 of the classical satellites have orbital poles
aligned (within $30^\circ$) with the normal of their spatially defined
plane. They suggested that the classical satellites may occupy a
rotationally-supported disc. \cite{li08} found that coherence in the
motions of the satellites may be due to group infall, whereby
satellites fall into the parent halo together and preserve their
common motion to the present day. A scenario whereby satellite
accretion is along surrounding filamentary structures suggests a link
between the angular momentum orientation of the satellite galaxies and
the host halo itself (\citealt{libeskind05}; \citealt{lovell10}). A
correlation between the orbital motion of satellite galaxies and the
spin of their parent light distribution was seen by \cite{azzaro06} in
a carefully selected sample of SDSS galaxies. However, this bias
towards co-rotation was not seen by \cite{hwang10} in a larger sample
of SDSS host galaxies. The authors find equal numbers of satellites in
prograde and retrograde orbits. Numerical simulation studies by \cite{warnick06}
and \cite{shaw06} find a bias towards satellites on co-rotating orbits
relative to the net spin of their host halo. However, both of these
studies base their conclusions on results derived from dark matter
only simulations, and they focus on cluster-sized
haloes. Observational estimates are frustrated by uncertainties
regarding the spin direction of the parent galaxies and contamination
by interlopers. On the other hand, theoretical work has solely focused
on dark matter only simulations --- the orientation of the satellites'
orbit with respect to the \textit{stellar} distribution is yet to be
tested.

Dark matter only simulations have been hugely influential in
developing our knowledge of the large scale structure of the
Universe. However, some of the potential shortcomings of the standard
model posed by observations of our own Milky Way galaxy are difficult
to reconcile within a simulation that does not include luminous
matter. Previous work has made use of semi-analytic models in order to
include the necessary baryonic processes into these cosmological
simulations (e.g. \citealt{white91}; \citealt{cole94};
\citealt{somerville99}; \citealt{baugh06}). Whilst these methods have
provided valuable insights into the effects of baryonic physics on
galaxy formation, their limited spatial information makes them
unsuitable to investigate the processes of interest in this paper. For example, a key
assumption in these semi-analytic methods is that the angular momentum of the disc
is aligned with the spin of the dark matter halo.

In recent years, increasingly realistic implementations of the hydrodynamic
evolution of the baryons have become possible within the framework of cosmological
simulations which are able to match a series of galaxy properties and
scaling relations (e.g. \citealt{gnedin04}; \citealt{governato07};
\citealt{agertz09}; \citealt{crain09}; \citealt{crain10}; \citealt{font10}). In
particular, these hydrodynamic cosmological simulations are able to
follow the changes in shapes and angular momenta of the dark matter
and baryons self-consistently. In this paper, we make use of the \textsc{gimic} suite
of simulations described in detail by \cite{crain09}. This is a
re-simulation of 5 cosmologically representative regions ($\sim
20h^{-1}$ Mpc in radius) from the \textit{Millennium simulation}
(\citealt{springel05b}). In these regions, \textsc{gimic} incorporates
baryonic physics and achieves higher resolution than the
\textit{Millennium simulation} as a whole.

We use these simulations to study the orbital properties of satellite
galaxies in late-type galaxies. In contrast to previous work, we probe
the dynamics of the satellites relative to their host's stellar
component as well as the unseen dark matter component. In \S2, we
describe the \textsc{gimic} suite of simulations in more detail and
outline the selection criteria for our sample of parent haloes. In
\S3, we discuss the relation between the galaxy and the dark matter halo in both spatial and velocity
space. \S4 focuses on the satellite galaxies associated with our
sample of haloes. We investigate their spatial and angular momentum
distribution relative to both the galaxy and dark matter
halo. In \S5, we briefly outline an application of our results to test
estimators of the parent halo mass. Finally, in \S6 we draw our main
conclusions.

\section{The numerical simulations}

In this section, we briefly describe the simulations we have used and
outline our methods for selecting parent galaxy haloes and their
associated satellite galaxies.

\subsection{GIMIC}

The Galaxies-Intergalactic Medium Interaction Calculation
(\textsc{gimic}) suite of simulations is described in detail in
\citet{crain09} (see also \citealt{crain10}).  It consists of a set of
hydrodynamical re-simulations of five nearly spherical regions ($\sim
20 h^{-1}$ Mpc in radius) extracted from the {\it Millennium
  Simulation} (\citealt{springel05b}.  The regions were selected to
have overdensities at $z=1.5$ that represent $(+2, +1, 0, -1, -2)
\sigma$, where $\sigma$ is the root-mean-square deviation from the
mean on this spatial scale.  The 5 spheres therefore encompass a wide
range of large-scale environments. In the present study, we select
systems with total `main halo' (i.e., the dominant subhalo in a
friends-of-friends group) masses similar to that of the Milky Way,
irrespective of the environment.  \cite{crain09} found that the
properties of systems of fixed main halo mass do not depend
significantly on the large scale environment (see, e.g., Fig.\ 8 of
that paper).

We present only a brief summary of the \textsc{gimic} simulations
here, and refer to Crain et al. (2009, 2010) for more detailed
descriptions. The cosmological parameters are the same as those in the
{\it Millennium Simulation} and correspond to a $\Lambda$CDM model
with $\Omega_{\rm m} = 0.25$, $\Omega_{\Lambda} = 0.75$, $\Omega_{\rm
  b} = 0.045$, $\sigma_{8} = 0.9$ (where $\sigma_{8}$ is the rms
amplitude of linear mass fluctuations on $8 h^{-1}$ Mpc scale at
$z=0$), $H_{0} = 100 h$ km s$^{-1}$ Mpc$^{-1}$, $h = 0.73$, $n_s=1$
(where $n_s$ is the spectral index of the primordial power spectrum).

The simulations were evolved to $z=0$ using the TreePM-SPH code
\textsc{gadget}, described in \citealt{springel05a}.
Subsequently, the \textsc{gadget} code has been substantially modified
to incorporate baryonic physics which includes:

\begin{itemize}
\item
a prescription for star formation outlined in \cite{schaye08} that is
designed to reproduce the observed Kennicutt-Schmidt law
(\citealt{kennicutt98});
\item
radiative gas cooling in the presence of a UV/X-Ray background (see
\citealt{haardt01}) which includes the contribution
of metal-line cooling (computed element-by element) (\citealt{wiersma09a});
\item
the timed release of 11 individual metals by both massive (Type II SNe
and stellar winds) and intermediate mass stars (Type IA SNE and
asymptotic giant branch stars) (\citealt{wiersma09b});
\item
a kinetic supernova feedback model (\citealt{dallavecchia08}) which can quench
star-formation in low mass haloes and pollute the IGM with metals.
\end{itemize}

In the present study we analyse the `intermediate' resolution
\textsc{gimic} simulations which have $8$ times better mass resolution
than the \textit{Millennium simulation}. These runs have a dark matter particle mass
$M_{\rm dm} \sim 5.30 \times 10^{7} h^{-1}$ M$_{\odot}$ and an initial
gas particle mass of $M_{\rm g} = 1.16 \times 10^{7} h^{-1}$
M$_{\odot}$. This implies that it is possible to resolve systems with masses similar to that of the classical dwarf
galaxies. The remainder of the
$500 (h^{-1}$ Mpc)$^{3}$ Millennium volume is modelled with dark
matter particles at much lower resolution to ensure the presence of
surrounding large scale structure is accurately accounted for. Dark
matter only runs of the \textsc{gimic} simulations are also available
(at the same resolution) which can be directly compared to the
hydrodynamic versions of the simulations. The $-2\sigma$ volume of the
`high' resolution \textsc{gimic} simulations has also been carried out
to redshift $z=0$. This higher resolution run has $8$ times better
mass resolution than the intermediate runs (and $64$ times better than
the \textit{Millennium simulation}). We check that the conclusions
of this paper are not subject to resolution effects by ensuring our
main results are unchanged in the higher resolution simulations (see
Appendix A).

Previous work utilising the \textsc{gimic} simulations has found
encouraging agreement with observational studies. The adopted star
formation and feedback prescription results in a good match to the star formation rate history of the universe (Crain et
al. 2009; see also \citealt{schaye10}), and also reproduces a number
of X-ray/optical scaling relations for normal disc galaxies
\citep{crain10}. \cite{font10} find that the stellar haloes of
their simulated Milky way-mass galaxies have luminosities and radial
density profiles in good agreement with observations. Furthermore,
McCarthy et al. (In prep) show that the simulated  $L^*$ disc galaxies
in \textsc{gimic} have realistic kinematics and sizes.

\subsection{Identification of Galaxies and Satellites}

\begin{table}
\begin{center}
\renewcommand{\tabcolsep}{0.12cm}
\renewcommand{\arraystretch}{0.005}
  \begin{tabular}{|l  l  l  l  l|}
    \hline 
    $\langle r_{200} \rangle (\mathrm{kpc})$ & $ \langle M_{200} \rangle (10^{12} M_\odot)$ &
    $ \langle M_{\star} \rangle (10^{10} M_\odot)$ &
    $\langle D/T \rangle $ & $\langle N_{\mathrm{sat}} \rangle$\\
    \\
    \hline
    224 & 1.4 & 8.0 & 0.7 & 9
    \\ 
    \hline
  \end{tabular}
  \caption{The average properties of our sample of (431) galaxy
    haloes. We give the median halo radius ($r_{200}$), the median
    total mass inside this radius ($M_{200}$), the median parent halo
    stellar mass ($M_{\star}$), the median disc-to-total stellar mass
    ratio ($D/T$) and the median number of satellite galaxies per
    parent halo ($N_{\mathrm{sat}}$).}
\label{tab:props}
\end{center}
\end{table}

Bound haloes are identified using the SUBFIND algorithm of
\cite{dolag09}, which extends the standard implementation of
\cite{springel01} by also including baryonic particles when identifying
self-bound substructures.  The main galaxy or parent halo is the
most massive subhalo belonging to a friends-of-friends system. The other self-bound
substructures of the system are then classified as satellite galaxies.

We define $r_{200}$ as the radius at which the mean enclosed density falls to
200 times the critical density ($200\rho_{\mathrm{crit}}$). We select
parent haloes with total mass within this radius in the range $5
\times 10^{11} < M_{200}/M_{\odot} < 5 \times 10^{12}$. Only haloes
with at least one associated satellite galaxy (or subhalo) other than
the main halo are included. \textsc{gimic} is a re-simulation and is
thus subject to edge effects at the boundaries of the selected
spherical regions. We discard any haloes located
at the boundary edges that are partially comprised of low resolution dark matter
particles. 

We select our sample of parent haloes according to the `relaxation'
criteria defined below, following the same reasoning as \cite{neto07}:

\medskip
\noindent
(1) Virial ratio: We compute the ratio $2K/|U|$, where $K$ and $U$ are
the total kinetic energy and total potential energy within
$r_{200}$. We adopt $2K/|U| < 1.35$ for the relaxed sample.

\medskip
\noindent
(2) Centre of mass displacement: The offset between the centre of mass
of the halo and position of the most bound particle (potential centre)
can be described by the normalised offset parameter,
$\epsilon=|r_c-r_{cm}|/r_{200}$ (\citealt{thomas01}). Relaxed haloes have $\epsilon <
0.07$.

\medskip
\noindent
(3) Mass in substructure: For a halo to be considered relaxed, we
require the fraction of mass in substructure within $r_{200}$ to be
$f_{\mathrm{sub}} < 0.1$ and the most massive satellite within
$2r_{200}$ to satisfy $M_{\mathrm{sat}}/M_{200} < 0.1$.

\medskip Under these criteria approximately 20\% of the haloes are
unrelaxed (c.f. \citealt{neto07}). There are a total of 624 relaxed
haloes within the given mass range. Most unrelaxed haloes are
recognised from the second and third constraints, whilst only a few
haloes are out of virial equilibrium. Note that since only a
relatively small fraction of haloes are unrelaxed, our conclusions are
generally valid.

\begin{figure*}
  \centering
  \includegraphics[width=10cm,height=10cm]{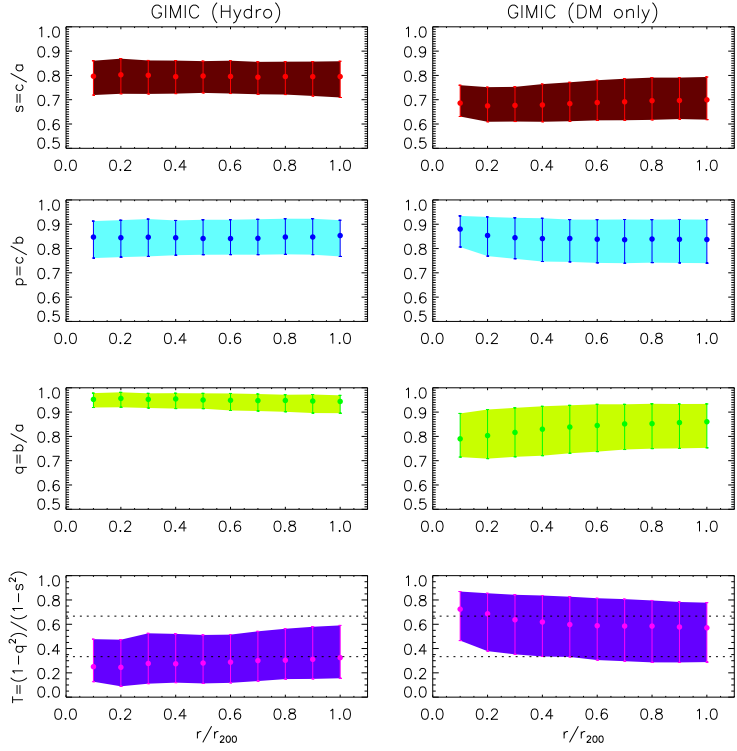}
  \caption{\small The radial dependence of the axial ratios $s=c/a$ (top
    panels), $p=c/b$ (second row), $q=b/a$ (third row) and the
    triaxiality parameter $T=(1-q^2)/(1-s^2)$ (bottom panels). The
    shaded regions show the range of values covered by 68\% of the
    distribution and the points give the median values.  The dotted
    lines indicate the regions of oblate ($T<1/3$), triaxial
    ($1/3<T<2/3$) and prolate haloes ($T>2/3$). The right hand panels
    are for the dark matter only counterparts of our sample. The halo
    shapes are rounder (and more oblate) in the hydro \textsc{gimic} simulations.}
  \label{fig:shapes}
\end{figure*}

The galaxies are assigned a morphological classification (i.e. disc
and spheroid-dominated types) based on their dynamics. A simple
two-component model is assumed: (i) a dispersion-supported spheroid,
and (ii) a rotationally supported disc. Details of this decomposition
into morphological types is given in \cite{crain10}. Fig. 1 in
\cite{crain10} shows the distribution of the disc-to-total stellar
mass ratios (D/T). 
In this work, we restrict our sample to late-type galaxies and exclude
obviously `elliptical' galaxies. Following the reasoning of
\cite{crain10}, we adopt a threshold of $D/T > 0.3$. We have checked
that our main results are unchanged if other cuts of $D/T$ are imposed
(e.g. $D/T > 0.2$ or $D/T > 0.4$). We note that haloes are
  selected based only on mass and $D/T$. This selection was not chosen
  to reproduce Milky Way (or M31) galaxy replicas and comparison with
  these two Local Group galaxies is made in the broadest sense.

Our final sample of haloes consists of 431 parent haloes and 4864
associated satellite galaxies. We summarise the properties of our
sample in Table \ref{tab:props}. The approximate (total) mass
  range of our sample of satellite galaxies is $10^8M_\odot < M_{\rm
    sat} < 10^{11}M_{\odot}$. Previous work using low resolution dark
matter only simulations have much larger samples of parent haloes
(e.g. \citealt{sales07}, who use thousands of parent haloes from the
\textit{Millennium simulation}). However, our sample size compares
favourably to more recent hydrodynamical simulations
(e.g. \citealt{libeskind07}, who had only 9 parent haloes).

In addition to our hydrodynamical suite of \textsc{gimic} simulations,
we have also the dark matter only runs.  We can match our sample of
haloes with their dark matter only counterpart. For each halo, we
consider dark matter only haloes with similar masses inside $r_{200}$
(within a factor of 2). We then compute the distances between the
positions of the dark matter only haloes and the position of the halo
in question. The dark matter only counterpart is then the closest (in
position) to the baryonic simulations version of the halo which also has a similar
mass.

\begin{figure*}
  \centering
  \includegraphics[width=10cm,height=10cm]{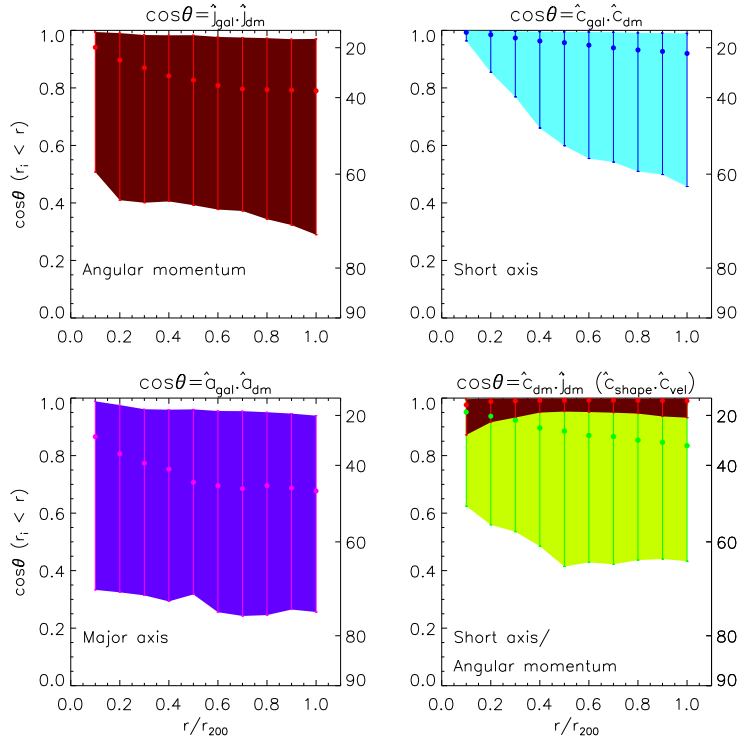}
  \caption{\small Median misalignment angles as a function of
    radius. Error bars and shaded regions show the values covered by
    68\% of the distribution (i.e. the $1\sigma$ dispersion). Top
    left: Misalignment angles between the angular momentum vector of
    the galaxy ($\hat{j}_{\rm gal}$) and the angular momentum
    vector of the dark matter halo ($\hat{j}_{\rm dm}$). Top right:
    Misalignment angle between the short axis of the galaxy
    ($\hat{c}_{\rm gal}$) and the short axis of the dark matter
    distribution ($\hat{c}_{\rm dm}$). Bottom left: Misalignment
    angles between the major axis of the galaxy ($\hat{a}_{\rm
      gal}$) and the major axis of the dark matter distribution
    ($\hat{a}_{\rm dm}$). Bottom right: Misalignment angles between
    the short axis and the angular momentum vector of the dark matter
    distribution (green shaded region). For comparison the
    misalignment between the short axis of the velocity anisotropy
    tensor ($\hat{c}_{\rm vel}$) and the short axis of the dark matter
    halo shape ($\hat{c}_{\rm shape}$) is shown by the red shaded
    region. There are strong alignments between the galaxy and
    the dark matter halo at small radii ($r \sim 0.1r_{200}$) but
    there can be significant misalignments at larger radii ($r \sim
    r_{200}$).}
  \label{fig:halo_align}
\end{figure*}

\begin{figure*}
  \centering
  \begin{minipage}{0.9\linewidth}
    \centering
    \includegraphics[width=12cm,height=5.5cm]{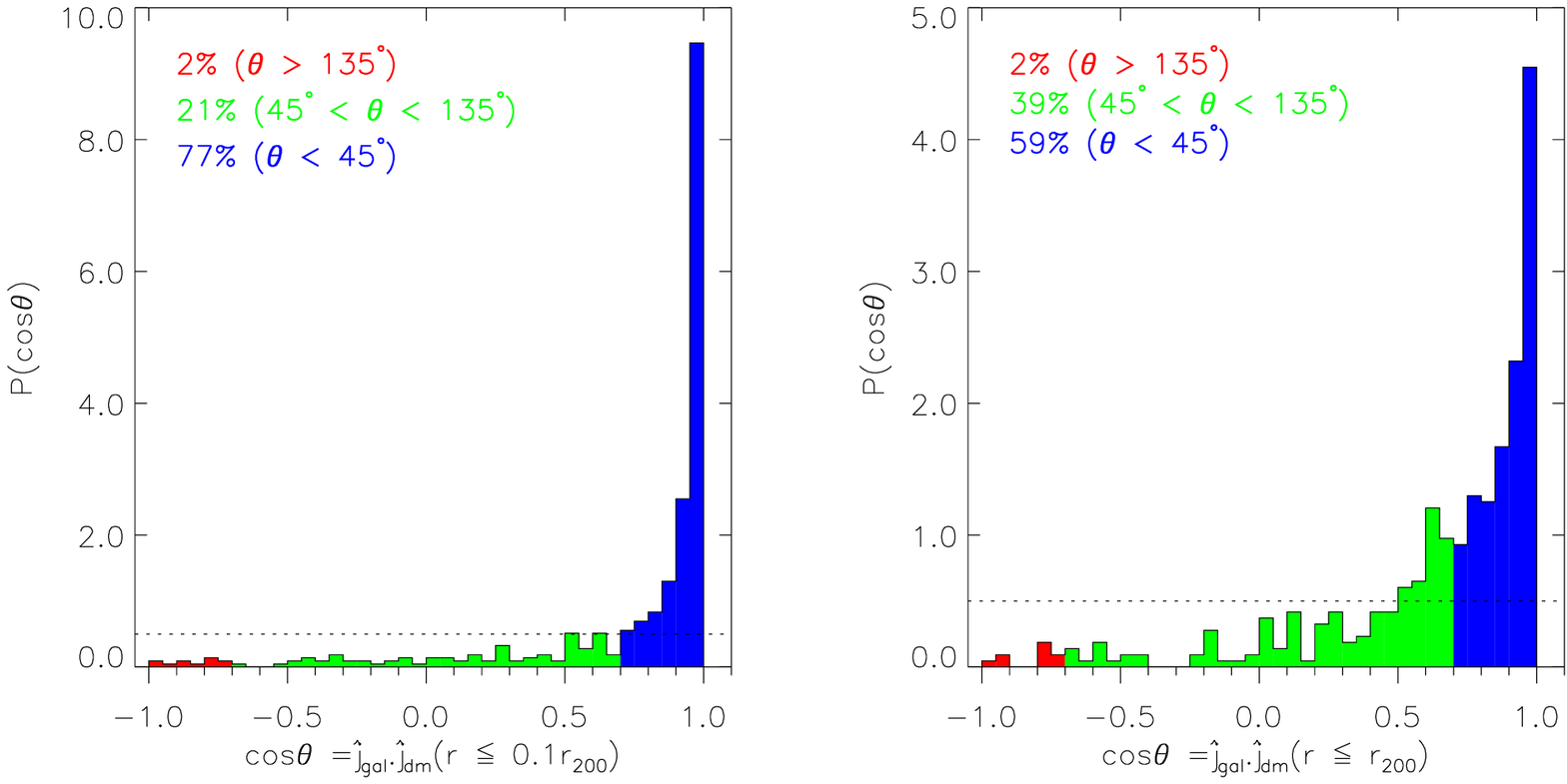}
  \end{minipage}
  \begin{minipage}{0.9\linewidth}
    \centering
    \includegraphics[width=12cm,height=5.5cm]{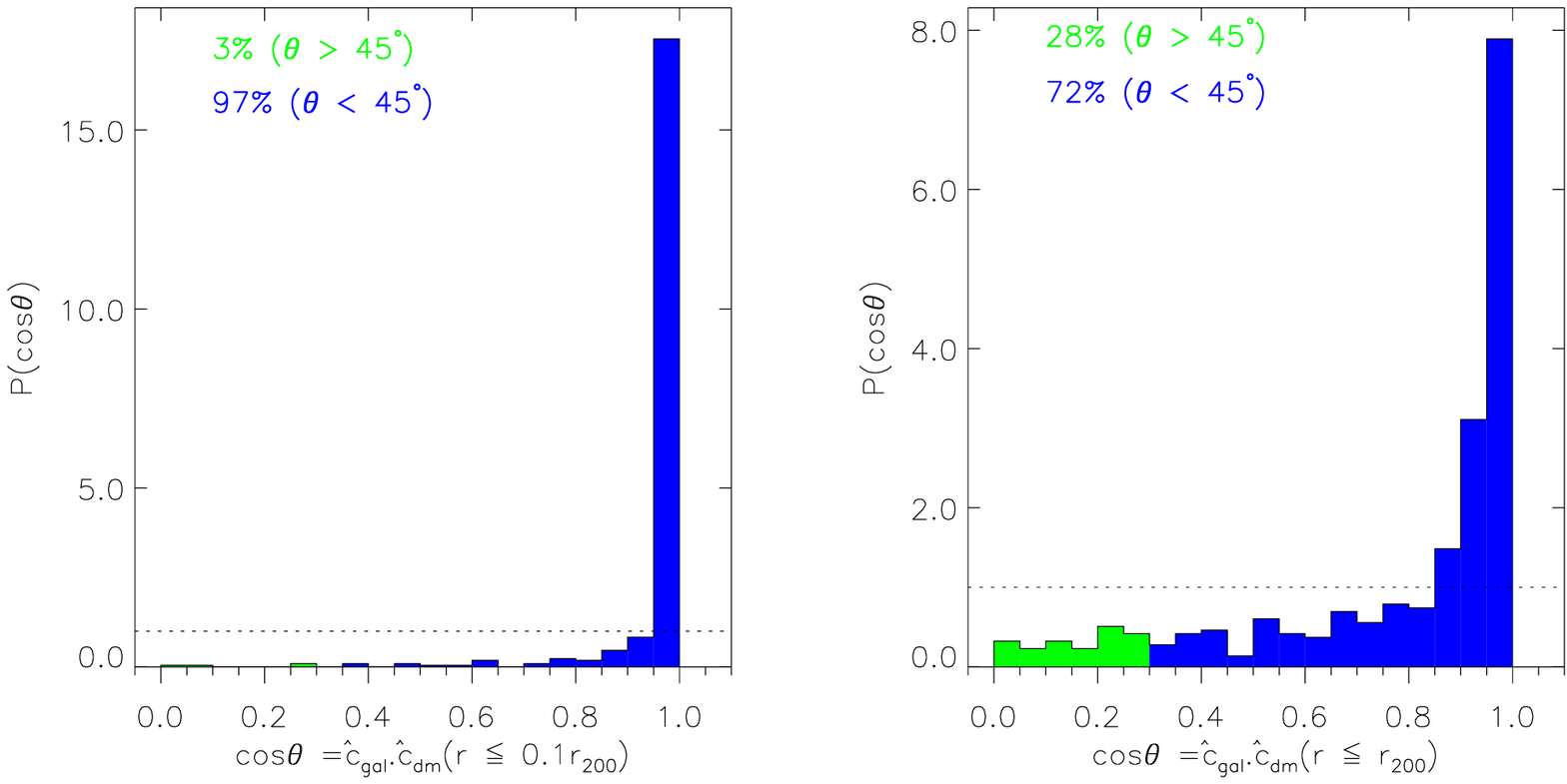}
  \end{minipage}
  \caption{\small Top: The distribution of misalignment angles between the
    angular momentum of the galaxy and the angular momentum of
    the dark matter halo for $r \le 0.1r_{200}$ (left panel) and $r
    \le r_{200}$ (right panel) respectively. Bottom: The distribution
    of misalignment angles between the short axis of the galaxy
    and the short axis of the dark matter halo for $r \le 0.1r_{200}$
    (left panel) and $r \le r_{200}$ (right panel) respectively. The
    dotted lines show a uniform distribution.}
  \label{fig:histogram}
\end{figure*}

\section{Halo Properties}
\label{sec:halo_props}

Here, we examine the shapes, the angular momenta and the
misalignments of the dark matter haloes of our simulated galaxies.

\subsection{Shapes}

Haloes are modelled as ellipsoids characterised by three axes, $a,b,c$
where $a \ge b \ge c$. The axial ratios $s=c/a$, $q=b/a$ and $p=c/b$
describe the three classes of ellipsoids: prolate ($a>b \approx c$),
oblate ($a \approx b > c$) and triaxial ($a>b>c$).  The shapes of dark
matter haloes are computed using the weighted (or reduced) second
moment tensor
\begin{equation}
\label{eq:inertia}
I_{ij}=\sum^{N(\le r)}_{n=1} \frac{x_{i,n}x_{j,n}}{r_{n}^2}
\end{equation}
where
\begin{equation}
r_n=\sqrt{x_{1,n}^{2}+\frac{x_{2,n}^{2}}{q^2}+\frac{x_{3,n}^{2}}{s^2}}.
\end{equation}
The advantage of this scheme is that every particle is given equal
weight, independent of radius. The orientation is defined by the
eigenvectors of the second moment tensor. The principal components of
the tensor give the axial ratios
\begin{equation}
q=\sqrt{\frac{I_{yy}}{I_{xx}}}, \, \, \, \, s=\sqrt{\frac{I_{zz}}{I_{xx}}}
\end{equation}
As the value of the elliptical radius, $r_{n}$, is not known in
advance (it depends on $s$ and $q$), the axial ratios are computed
using an iterative algorithm (\citealt{dubinski91}). In Figure \ref{fig:shapes}, we show the radial
dependence of the three axis ratios. The bottom panel gives the radial
behaviour of the triaxiality parameter (\citealt{franx91}) defined by
\begin{equation}
T=\frac{1-(b/a)^2}{1-(c/a)^2}=\frac{1-q^2}{1-s^2}.
\end{equation}
Oblate, triaxial and prolate haloes have triaxiality parameters of
$T<1/3$, $1/3<T<2/3$ and $T>2/3$ respectively. The right hand panels
of Fig. \ref{fig:shapes} give the axis ratios for the dark matter only
counterparts of our sample. The shaded regions show the values covered
by 68\% of the distribution and the points denote the median values.

The dark matter haloes in our sample are close to spherical and
are slightly more oblate in the inner regions. This is in stark
contrast to the haloes in the dark matter only simulations which
become more prolate towards the centre. The inclusion of baryonic
physics affects the shape of the halo substantially in the inner
regions, but also has a significant effect throughout the halo.

Our results broadly agree with the findings of previous studies but
direct comparisons are difficult as many authors have concentrated on
cluster-sized haloes or have used simulations where the assembly of
the central galaxy is not modelled
self-consistently. \cite{dubinski94}, for example, simulates
dissipative infall by growing a central mass concentration inside a
triaxial dark matter halo and find that a steeper potential leads to
rounder and more oblate dark matter halo shapes ($\Delta(c/a) \sim
0.1$). \cite{kazantzidis04} study cluster-sized haloes using
gas dynamical simulations and find they are significantly rounder in
the inner regions ($\Delta(c/a) \sim 0.2$) but the changes are
radially dependent and are almost negligible at the virial
radius. \cite{abadi10} employ galaxy models which include radiative
gas cooling but neglect the contribution from stellar feedback. They
compute equipotential axial ratios and find roughly constant
flattening of $\langle c/a \rangle \sim 0.85$, which is significantly
rounder than their dark matter only runs. By contrast, we have
characterised the halo shape by the density of the dark matter,
which is always flatter than the equipotential surfaces. Compared to
\cite{abadi10}, we do not find such a significant change in sphericity
relative to the dark matter only simulations. This is presumably
because the \textsc{gimic} simulations include stellar feedback and do
not suffer from strong overcooling (see e.g. \citealt{duffy10}). As
indicated in Table \ref{tab:props}, the mean stellar mass fraction of
our simulated disc galaxies is $\sim 0.057$. This corresponds to a
baryon conversion efficiency ($M_*/M_{\rm halo} \times
\Omega_m/\Omega_b$) of $\approx 30\%$, which is only a factor of
$\sim1.5$ larger than that inferred recently by \cite{guo10} by
matching the observed stellar mass function from the most recent SDSS
data release to the halo mass function derived from the
\textit{Millennium} and \textit{Millennium-II} simulations. 

Unfortunately, the evidence on the shape of the Milky Way's dark halo
is far from clear-cut.  Many authors have argued that the coherence of
the Sagittarius tidal stream may constrain the halo shape. However,
this line of enquiry has concluded that the halo may be almost
spherical (\citealt{ibata01}; \citealt{fellhauer06}),
oblate~(\citealt{johnston05}), prolate~(\citealt{helmi04}) or
triaxial~(\citealt{law10}). This variety of results strongly suggests
that halo shape is not the primary factor determining the complex
morphology of the Sagittarius stream. \cite{smith09} argued that the
spherical alignment of the velocity ellipsoid of SDSS halo subdwarf
stars implied that the gravitational potential and hence the dark halo
was nearly spherical. There is however contradictory evidence from
studies of the flaring of the HI gas layer by \cite{olling00}, who
found a highly flattened oblate ($q \approx 0.3$) dark halo for the
Milky Way.

The dark matter halo shapes of external galaxies are often estimated
using galaxy-galaxy weak lensing studies. \cite{hoekstra04} and \cite{parker07} both
found an average axis ratio of $\sim 0.7$ from their studies using the
Red-Sequence cluster survey and the CFHT legacy survey,
respectively. However, \cite{mandelbaum06} found no evidence for halo
ellipticity in their study using SDSS data.

\subsection{Angular Momentum and Shape}

The (cumulative) specific angular momentum vector for the dark matter particles,
$\mathbf{j_{dm}}$, and the stellar component, $\mathbf{j_{gal}}$, are
defined as
\begin{equation}
\label{eq:ang_mom}
\mathbf{j(\le r)}=\frac{1}{M( \le r)} \sum^{N(\le r)}_{n=1}m_n \mathbf{x_n} \times
\mathbf{v_n}
\end{equation}
where $\mathbf{x_n}$ and $\mathbf{v_n}$ are the position and velocity
vectors of particle $n$ relative to the halo centre and the
centre-of-mass velocity.  We compute $\mathbf{j_{gal}}$ for $r \leq
0.1r_{200}$ (using only the star particles) to characterise the inner
galaxy.  The stress or velocity dispersion tensor of the dark matter
distribution is
\begin{equation}
\label{eq:aniso_ten}
\Pi_{ij}=\sum^{N \le r}_{n=1} v_{i,n}v_{j,n}.
\end{equation}
We can diagonalise this tensor to find the eigenvectors and
eigenvalues which define the principal velocity anisotropy axes
($a_{\mathrm{vel}},b_{\mathrm{vel}},c_{\mathrm{vel}}$).

In Fig. \ref{fig:halo_align}, we show the median misalignment angles
between the galaxy and the dark matter halo as a function of
radius. The top left panel shows the misalignment in angular
momentum, the top right the misalignment in shape. Note that the
angular momentum or shape of the galaxy is always calculated for
$r \le 0.1r_{200}$, whereas the computation of the dark matter halo
properties varies with radius.  There is strong alignment of angular
momentum vectors in the inner regions of the halo, but the median
misalignment gradually increases to $\approx 40^\circ$ at $r \sim
r_{200}$. This is in good agreement with \cite{bett10} who find
a median misalignment of $\sim 30^\circ$ for a smaller sample of $\sim 50$ haloes. There is also strong
alignment between the short axis of the galaxy and the short axis of the dark
matter halo where the median misalignment grows to $\approx 20^\circ$
at $r \sim r_{200}$.  In the bottom left hand panel, we see that the
alignment between the major axes of the galaxy and dark matter
halo is poorer and there is significant scatter. This is not
surprising as for disc-like configurations the major axis is poorly defined
(i.e. $q=b/a \approx 1$). The bottom right hand panel shows the
alignment between the short axis and the angular momentum vector of
the dark matter halo as a function of radius (shaded green). We also
show as the shaded red region the alignment between the short axis of
the dark matter halo shape and the short axis of the velocity
dispersion tensor. There is very strong alignment between the shape
and velocity dispersion. This reflects the fact that the dark matter halo shape
is supported by internal velocities rather than net rotation (\citealt{frenk88}).

To avoid any ambiguity in the definition of the short axes or angular
momentum vectors of the dark matter haloes, \cite{bett10} imposed
constraints on their shapes and net angular momentum ($s < 0.81$,
$\mathrm{log}_{10} j(\leq r_{200})/\sqrt{GM_{200}r_{200}} \geq
-1.44$). With these restrictions, we see stronger alignments in
general between the galaxy and dark matter halo. This is not
surprising, as we have found that the dark matter haloes are close to
spherical and are mainly dispersion, rather than rotationally,
supported. However, we choose to include all of our haloes to avoid
selection biases ($only \sim 40 \%$ of our sample satisfy the
constraints), but check that our main results are not significantly
altered by discarding dark matter haloes that have little net rotation
and are close to spherical.

Further insight into the misalignment between the galaxy and dark
matter halo can be gained by examining the distribution of these
misalignment angles at different radii. In Fig. \ref{fig:histogram},
we show the distributions of misalignments between the galaxy and dark
matter halo angular momentum vectors (top panels) and short axes
(bottom panels) both for $r\le 0.1r_{200}$ (left panels) and for $r
\le r_{200}$ (right panels). As we already saw in
Fig. \ref{fig:halo_align}, there is much stronger alignment in the
inner regions of the halo. However, when the shape of the dark matter
halo is computed for $r \le r_{200}$, the short axis of the inner
galaxy is misaligned from the short axis of the dark matter halo by
$\theta > 45^\circ$ in approximately $30\%$ of the systems. The
orientation of the angular momentum vector of the galaxy is aligned
almost perpendicularly ($45^\circ < \theta < 135^\circ$) to the net
spin of the dark matter halo in approximately $40\%$ of the systems
and $2\%$ are even anti-aligned. Our results are in good agreement
with \cite{bailin05a} who analyze the shape and internal alignment of
the dark matter halos of seven high-resolution cosmological disk
galaxy formation simulations. The authors find that while the minor
axis of the inner dark matter halo ($ r \le 0.1r_{200}$) is well
aligned with the disc axis, the outer regions of the halo can be
substantially misaligned.

The misalignments between the galaxy and the dark matter halo have
potential implications for galaxy-galaxy weak lensing studies. Such
studies attempt to deduce the weak cosmological signal from the
observed galaxy ellipticity correlation. Typically the influence of
any intrinsic signal on the observed ellipticity correlation is
neglected. However, correlations between halo shape and the density
field are expected to arise through tidal torques operating during
galaxy formation (e.g. \citealt{heavens88}). More recent work has
attempted to quantify the potential sources of lensing contamination
caused by these intrinsic alignments by examining dark matter only
cosmological simulations (e.g. \citealt{heymans06}). Any misalignments
between the galaxy and the dark matter halo will weaken the expected
intrinsic signal deduced from dark matter only simulations.

What causes these misalignments between the galaxy and the outer dark
matter halo? According to tidal torque theory (e.g. \citealt{white84};
\citealt{fall80}), the angular momentum of the galaxy and the dark
matter halo are initially very well aligned. However, the outer halo
continues to accrete material, which can alter its shape and/or its
net angular momentum. Hence, whilst we see strong alignment in the
inner regions of the halo there can be significant misalignments in
the outer parts.  This has important implications for the satellite
populations of these haloes. The majority of the satellite galaxies
are located in the outer regions of the halo. It is to this topic that
we now turn.

\begin{figure}
  \centering
  \begin{minipage}{\linewidth}
    \centering
    \includegraphics[width=7cm,height=6.3cm]{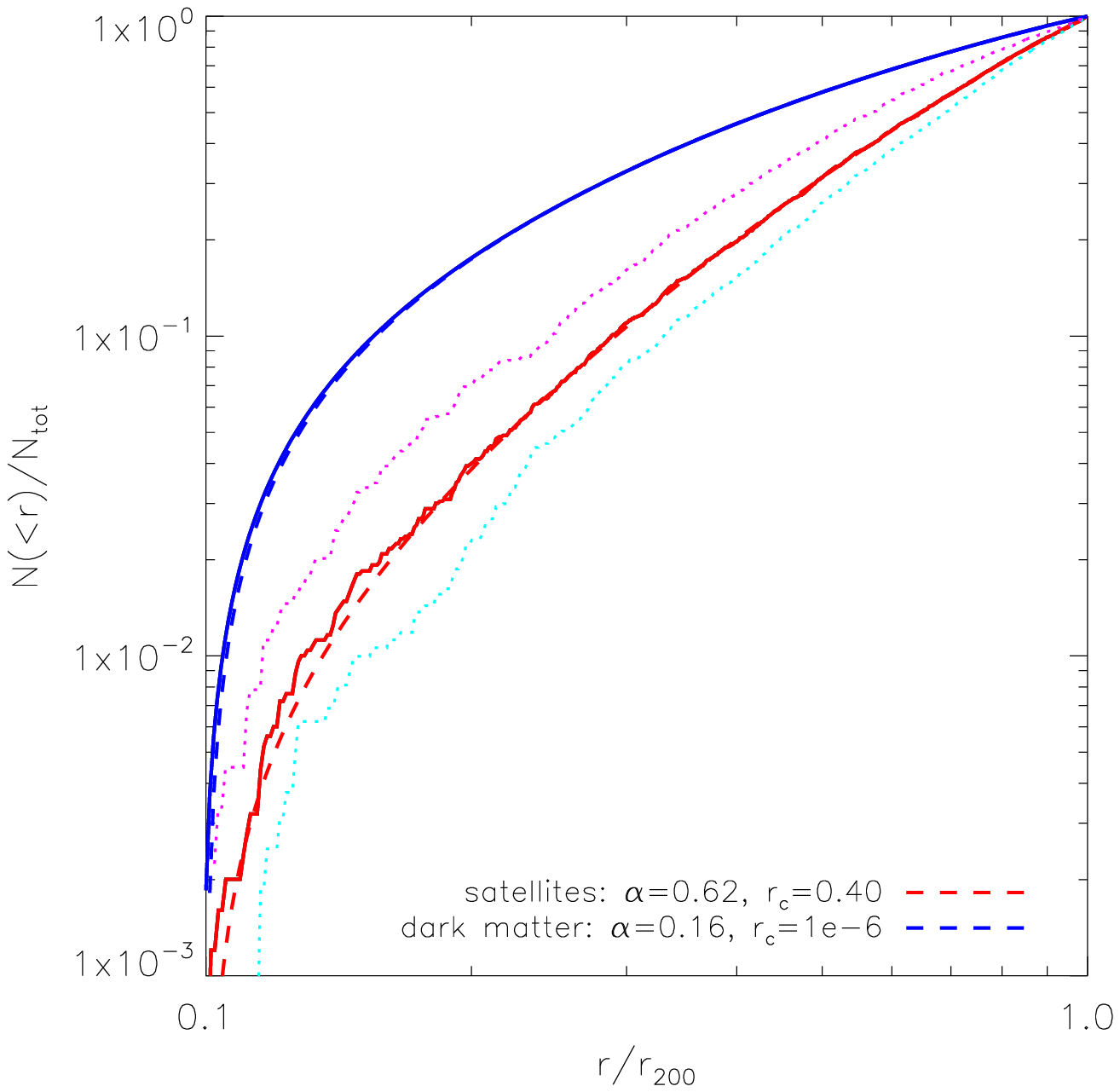}
  \end{minipage}
  \begin{minipage}{\linewidth}
    \centering
    \includegraphics[width=7cm,height=6.3cm]{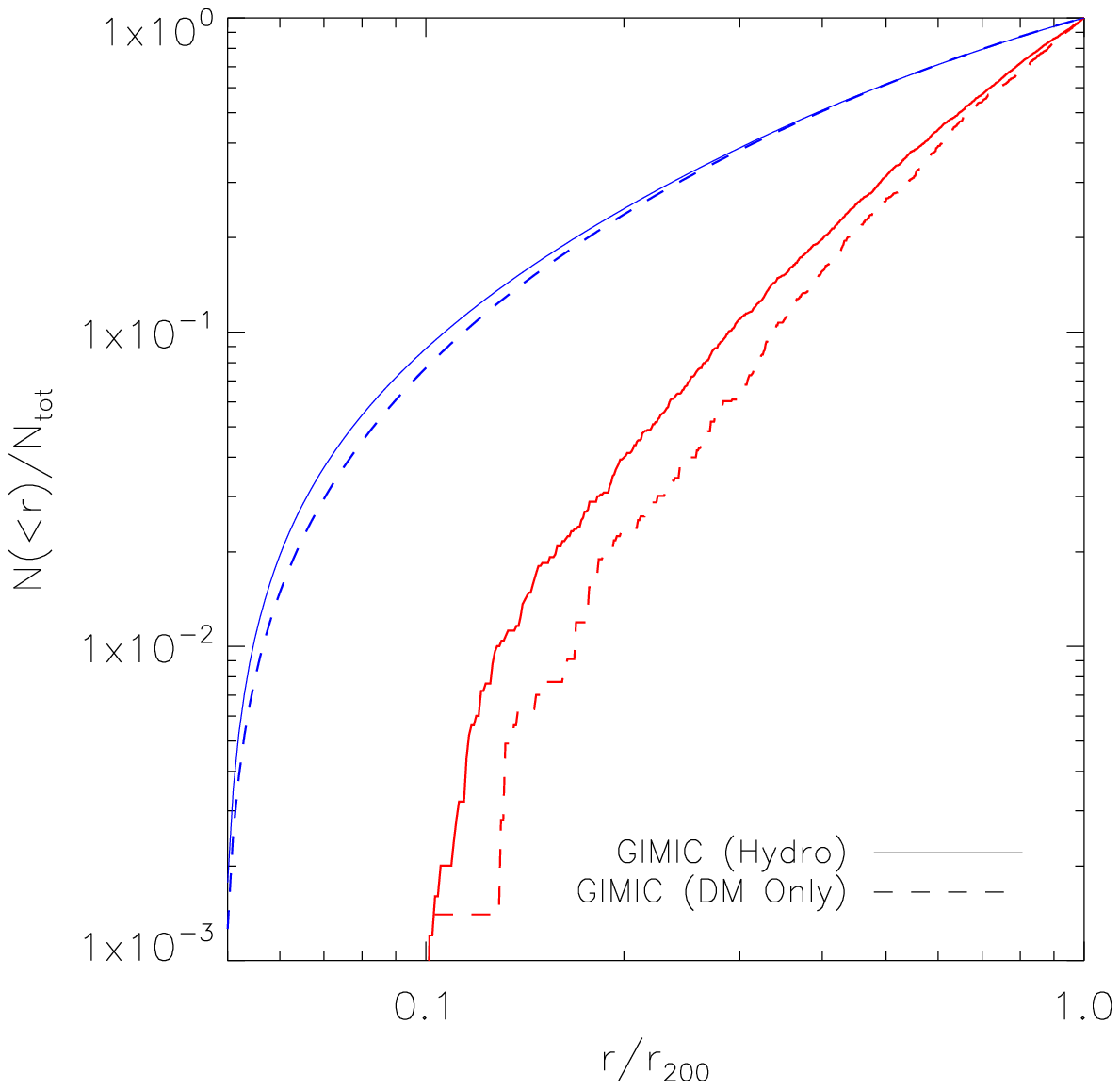}
  \end{minipage}
  \begin{minipage}{\linewidth}
    \centering
    \includegraphics[width=7cm,height=6.3cm]{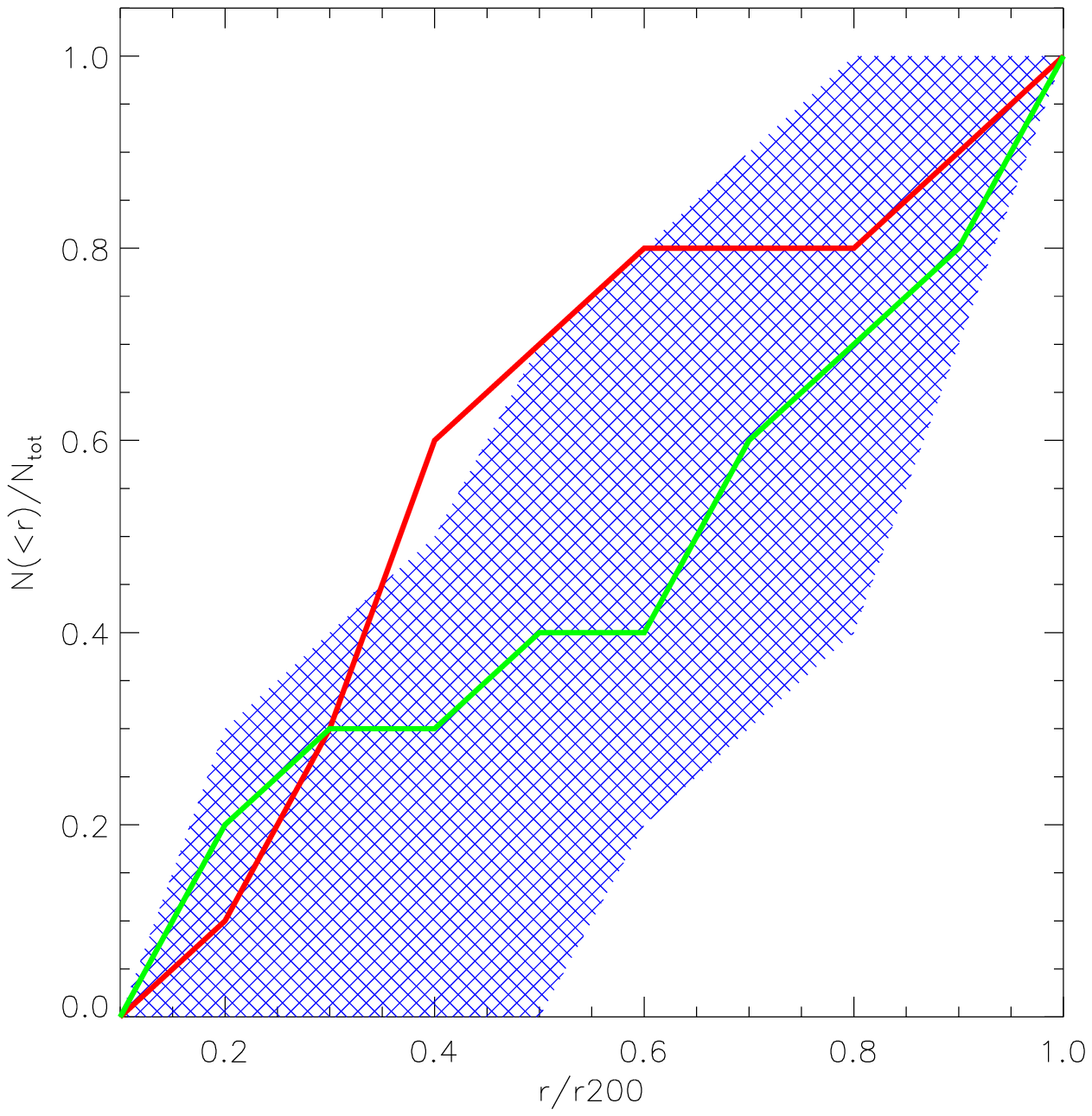}
  \end{minipage}
   \caption{\small Top panel: The cumulative number distribution for
     satellites (full red line) and dark matter particles (full blue line)
     within $r_{200}$. Fits to Einasto profiles are given by the red
     and blue dashed lines. The dotted magenta and cyan lines show
     the cumulative number distribution for satellites with stellar
     mass fractions, $M^*_{\rm sat}/M_{\rm sat}$, greater
     than 2$\%$ and less than 2$\%$ respectively. Middle
     panel: The red and blue lines give the radial profiles of the
     satellites and dark matter particles respectively. The dashed
     lines show the profiles for the dark matter only counterparts of
     our sample of haloes. Bottom panel: The cumulative number
     distribution for the ten brightest satellites of individual
     systems. The blue hashed region encompasses the scatter of the
     profiles for the $\sim 80$ systems with ten or more satellites within
     $r_{200}$. The red and green lines give the profiles for the
     classical Milky Way satellites and the ten brightest M31
     satellites (within $r_{200}$) respectively.}
  \label{fig:dens}
\end{figure}

\section{Satellite Galaxies}
\label{sec:sats}
Here, we study the spatial and velocity distributions of the satellite
systems of our simulated galaxies, focusing on their alignment (in
positional and velocity space) with the parent dark matter halo and
galaxy. 

In Figure \ref{fig:dens}, we show the cumulative number distribution
of satellite galaxies (red) and dark matter (blue) within
$r_{200}$. We have stacked all of the satellite galaxies by
normalising their radial distances from the parent halo centre by
$r_{200}$. The density profiles for both the dark matter particles and
satellite galaxies can be described (\citealt{navarro04}) by
\begin{equation}
\mathrm{ln}\rho(r) \propto -(2/\alpha)[(r/r_c)^\alpha-1].
\end{equation}
This density profile was first introduced by Einasto (see
e.g. \citealt{einasto89}) and is mathematically equivalent to the
Sersic profile that is often used to describe the projected density
profile of galaxies. Low values of the power-law slope ($\alpha$)
describe a `cuspy' density profile. In agreement with previous work
(e.g. \citealt{navarro04}; \citealt{gao08}), we find a dark matter
density slope of $\alpha_{\mathrm{dm}} \sim 0.2$. In contrast, the
satellite distribution favours larger values of $\alpha$ and has a
`cored' profile. As found many times before (e.g. \citealt{zentner05};
\citealt{libeskind05}; \citealt{ludlow09}), there is an obvious
spatial bias between the dark matter and the satellite
galaxy population.

Fig. \ref{fig:dens} also shows the cumulative number of satellites
with stellar mass fractions greater than 2$\%$ and less than 2$\%$
by the magenta and cyan dotted lines respectively\footnote{We verified
  that our results are not substantially affected if we choose
  fractional stellar mass limits of $1.5$ or $2.5\%$} . Satellites
with higher stellar mass fractions are more centrally located. These
are generally the most massive satellites which have spiralled into
the central regions of the halo during the course of several
pericentric passages. The stellar component of satellite galaxies is
more centrally located than the dark matter distribution. Whilst the
dark matter `envelope' is more effectively tidally stripped, the
stellar component remains relatively undisturbed. \cite{libeskind10}
showed that the increased density in the central regions of a subhalo
due to the collapse of baryons leads to a reduced mass loss relative
to an equivalent dark matter only satellite (see also
\citealt{maccio06}). Dynamical friction is thus more effective and
satellites sink closer into the parent halo. This leads to a more
centrally concentrated distribution of satellite galaxies when baryons
are included. By comparison with our dark matter only simulations, we
also see a more substantial inner radial bias when baryons are
included (see bottom panel of Fig. \ref{fig:dens}). Note that this
effect is not as pronounced for the lower mass satellites, which are
less affected by dynamical friction.

The bottom panel of Fig.~\ref{fig:dens} shows the cumulative number
distribution of the ten brightest (i.e. highest stellar-mass)
satellites within $r_{200}$ for individual systems. Approximately 80
haloes have at least ten satellites within $r_{200}$. We limit
ourselves to the ten brightest satellites to enable comparison with
the data on the Milky Way and M31. The positional data for the
  Milky Way and M31 satellites is taken from tables 3 and 5 of
  \cite{deason11} (see references therein). Although the top panel of
Fig.\ref{fig:dens} gives no indication of the scatter in the
simulations, the bottom panel explicitly shows the system-to-system
variation (encompassed by the blue hatched region). The solid red and
green lines gives the profiles for the classical Milky Way satellites
and the ten brightest M31 satellites (within $r_{200}$). We use the
$r_{200}$ values for the Milky Way and M31 recently estimated by
\cite{guo10} of approximately $250$ kpc and $290$ kpc
respectively\footnote{These were computed using the estimated halo
  masses given in Table 1. of \cite{guo10} and then their eqn (1) was
  used to estimate $r_{200}$}. The profiles for both M31 and the Milky
Way lie within the scatter of the simulations. In particular, there is
very good agreement with M31 for which (owing to our external view)
there are less biases in the observed sample. Note that \cite{font10}
show in their Fig. 1 that the luminosity function of the satellite
galaxies in \textsc{gimic} are also in good agreement with the
satellite systems of Local Group galaxies.

\begin{figure*}
  \centering
  \includegraphics[width=12cm,height=6cm]{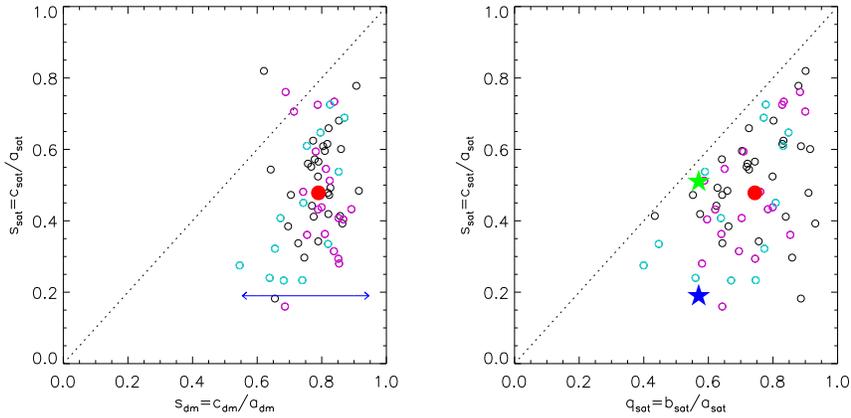}
  \caption{\small Left panel: The shape of the satellite distribution and underlying dark
    matter haloes as defined by the sphericity, $s=c/a$. Only haloes
    with ten or more satellites within $r_{200}$ are considered and
    the shape is computed for the ten highest stellar mass satellites. The
    red dot indicates the median values whilst the blue line indicates
    the approximate values for the Milky Way (the blue arrows denote
    the uncertainty in the sphericity of the dark matter halo of the
    Milky Way). The points are colour coded according to the alignment
    between the system of satellites and the galaxy. Cyan,
    magenta and black points are for satellite systems where
    $\hat{c}_{\rm gal}.\hat{c}_{\rm sat} > 0.75$, $\hat{c}_{\rm
      gal}.\hat{c}_{\rm sat} < 0.25$ and  $0.25 < \hat{c}_{\rm
      gal}.\hat{c}_{\rm sat} < 0.75$, respectively. The satellite
    galaxies generally have a more flattened spatial distribution than
    the dark matter. Right panel:
    The axial ratios of the satellite distribution. The blue and green
    stars
    indicate the axial ratios for the classical Milky Way satellites
    and the ten brightest M31 satellites (within $r_{200}$).}
  \label{fig:sat_shape}
\end{figure*}

\begin{figure*}
  \centering
  \includegraphics[width=12cm,height=6cm]{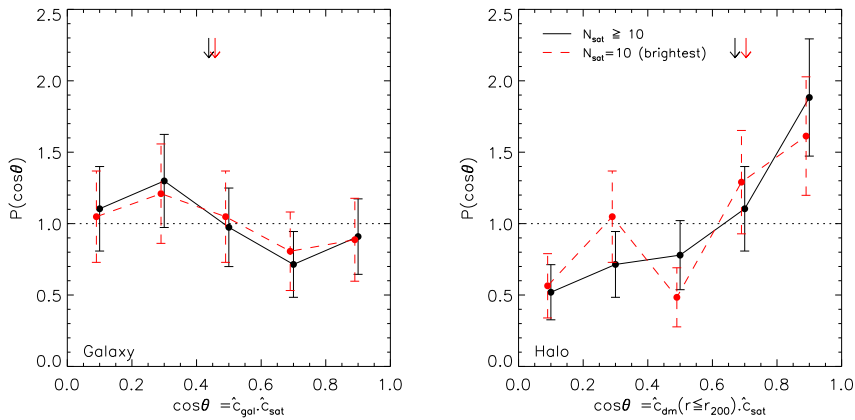}
  
   \caption{\small The orientation of the short axes of the satellite
    distribution relative to the short axes of the parent galaxy and dark matter halo (defined within $r_{200}$). The black lines are for all systems of satellites
    with ten or more satellites within $r_{200}$. The red dashed lines
    give the distributions when only the ten brightest satellites
    within $r_{200}$ are used to compute the shapes of the satellite
    distribution. Downward pointing arrows denote the median of the
    distributions and the dotted lines indicate a uniform
    distribution. The error bars denote Poisson uncertainties. Whilst
    there is strong alignment between the spatial distribution of the
    satellite galaxies and the dark matter halo, the distribution with
    respect to the galaxy is consistent with isotropy.}
   \label{fig:sats_shape_dmgal}
\end{figure*}

\subsection{Anisotropic Distributions}

\label{sec:sat_aniso}
Here, we first restrict attention to host galaxies which have ten or more
associated satellites within $r_{200}$ (c.f. Libeskind et al. 2007,
2009).  For comparison with the classical satellites of the Milky
Way (and M31), we consider the ten brightest (i.e. highest stellar mass)
satellites in each system. We compute the shape
($a_{\mathrm{sat}},b_{\mathrm{sat}},c_{\mathrm{sat}}$) of the
satellite galaxy distribution within $r_{200}$ of each host galaxy by
diagonalising the second moment tensor defined as
\begin{equation}
I^*_{ij}=\sum^{N(\le r)}_{n=1} x_{i,n}x_{j,n}.
\end{equation}
This is used in preference to the reduced inertia of
eqn~(\ref{eq:inertia}) which requires an iterative algorithm to
discard outliers until convergence is achieved. For systems of
satellites with a small number of data points, this is undesirable.

In the left hand panel of Fig. \ref{fig:sat_shape}, we show that the
satellite distribution is generally more flattened than the underlying
dark matter distribution. The red dot indicates the median axial ratio
values for the satellites and dark matter ($\langle s_{\mathrm{sat}}
\rangle \sim 0.5$, $\langle s_{\mathrm{dm}} \rangle \sim 0.8$). We
show with the blue arrow the range of values for the flattening of the
Milky Way dark halo given in the literature. We suggest the difference
in flattening may be understood by considering when the dark matter
was accreted relative to the satellite galaxies. Present-day
satellites are the surviving population, and have have been in orbit
for much less time than the dark matter (see Section
  \ref{sec:acc}). As far as the satellites are concerned, the
potential of the halo has largely been static since accretion.  The
dark matter, by contrast, has undergone relaxation and phase
mixing. Even though both the satellites and dark matter are accreted
anisotropically, this will be reflected to a greater degree in the
spatial distribution of the satellite galaxies rather than the dark
matter halo. The right hand panel of Fig. \ref{fig:sat_shape} shows
the distribution of axial ratios for the satellite systems.  The blue
and green stars show the flattening for both the classical Milky Way
satellites and the ten brightest M31 satellites computed in the same
way as the simulated satellites. The classical Milky Way satellites
have a highly flattened distribution which is consistent (within
$2\sigma$) with the median value derived from the simulations whilst
the M31 satellites have a less flattened configuration and lie closer
to the median $s=c/a$ value. The points are coloured according to the
alignment of the satellite distribution with the galaxy (see below and
Fig. \ref{fig:sats_shape_dmgal}). Polar ($\hat{c}_{\rm
  gal}.\hat{c}_{\rm sat} < 0.25$), planar ($\hat{c}_{\rm
  gal}.\hat{c}_{\rm sat} > 0.75$) and `in-between' ($0.25 <
\hat{c}_{\rm gal}.\hat{c}_{\rm sat} < 0.75$) alignments are given by
the magenta, cyan and black points respectively. There is no obvious
correlation between the orientation of the satellite distribution and
their flattening. Neither does there seem to be a bias towards a
particular orientation, as we discuss below.

\begin{table}
\begin{center}
\renewcommand{\tabcolsep}{0.12cm}
  \begin{tabular}{|l  c  c  c|}
    \hline 
     $D/T$ &$> 0.3$ & $> 0.5$ & $> 0.7$\\
    \hline
    $N_{\mathrm{tot}}$ & 431 & 328 & 192 \\
    $N(N_{\mathrm{sat}} \ge 10)$ & 86 & 68 & 51 \\
    $N(N_{\mathrm{sat}} \ge 10, \theta > 80^\circ)$ & 20 & 16 & 15
    \\
    $N(N_{\mathrm{sat}} \ge 10, \theta > 80^\circ, M^\star \sim M_{\rm
      MW})$ & 15 & 13 & 12 \\
    \hline
  \end{tabular}
  \caption{\small The number of parent haloes in the mass range
    $5\times 10^{11} < M_{200} < 5 \times 10^{12}$ for different $D/T$
    cuts. We give the total number of parent haloes, the number of
    haloes with at least 10 satellites within $r_{200}$, the number of
    satellite systems inclined by more than $80^\circ$ to the galaxy
    and the number of these systems which have similar stellar masses
    $M^\star$ (within a factor of 2) to the Milky Way and M31 ($\sim
    M_{\rm MW}$). Our results are robust to more restrictive $D/T$
    cuts. Approximately $20\%$ of satellite systems (with at least 10
    members) are aligned perpendicularly to their parent galaxy.}
\label{tab:angles}
\end{center}
\end{table}

\begin{figure}
   \centering
   \includegraphics[width=7cm,height=7cm]{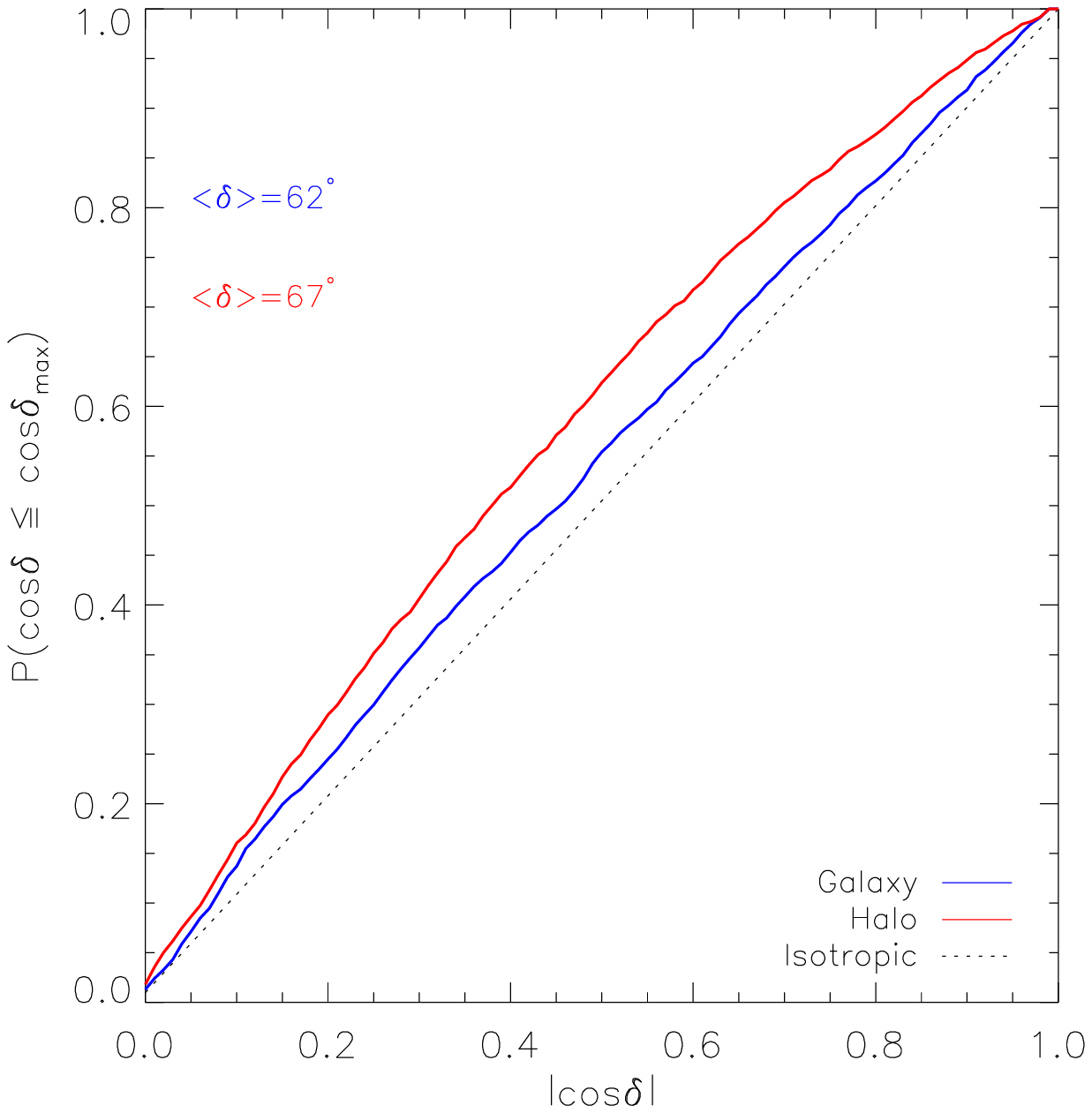}
   \caption{\small The cumulative probability distribution for
    the location of satellites relative to the host minor axis. Here,
    we consider \textit{all} satellites within $r_{200}$ and are not
    restricted to those systems with large numbers of
    satellites. $\delta$ is defined from the minor axis (i.e. $\delta =0$
    is polar alignment, $\delta=90^\circ$ is planar alignment). The blue
    distribution is relative to the short axis of the galaxy (or
    disc) and the red distribution is relative to the short axis of
    the dark matter halo (defined for $r \leq r_{200}$). There is a
    preference for planar alignment relative to the dark matter halo
    but there is a much weaker correlation with the galaxy.}
  \label{fig:sats_shape_dmgal2}
\end{figure}

We show the distribution of alignments between the short axes of the
satellite systems and the short axes of the dark matter halo (for $r
\le r_{200}$) and the galaxy of their parent haloes in
Fig. \ref{fig:sats_shape_dmgal}. The orientation of each satellite
system with respect to its host dark matter distribution and galaxy
are shown in the right and left panels respectively. The satellite
distribution preferentially aligns in a plane perpendicular to the
short axis of the dark matter distribution. However, the satellites
show no preferential alignment relative to the galaxy. There are a
significant number of systems where the satellite distribution is
aligned in a plane perpendicular to the disc ($20 \%$ of the systems
have $\cos \theta < 0.2$ or $\theta > 80^\circ$). Thus the alignment
of the Milky Way satellites perpendicular to the disc plane is not
inconsistent with the results we present here.  The red dashed lines
give the distributions when only the ten brightest satellites within
$r_{200}$ are used to compute the shapes of the satellite
distribution. Although there is substantial uncertainty caused by
small number statistics, we can see that by restricting ourselves to
the same number of satellites as the observational sample (and naively
ignoring selection biases), the apparent distribution of Milky Way
satellites is consistent with the simulations.

In Table \ref{tab:angles}, we give the number of satellite systems
which have similar (perpendicular) orientations as the Milky Way
Galaxy. Our results are robust to more restrictive $D/T$ cuts. In
fact, the fraction of systems with almost perpendicular satellite
alignments is slightly higher for parent galaxies with $D/T >
0.7$. Our total mass range ($5\times 10^{11} < M_{200}/M_{\odot} <
5\times 10^{12}$) is broadly coincident with the masses of the Milky
Way and M31. We can further refine our sample selection by using
haloes with stellar masses similar to these local galaxies. In the
bottom row of Table \ref{tab:angles}, we give the number of satellite
systems with polar alignments which also have parent stellar masses
within a factor of 2 of the observed estimates for the Milky Way and
M31 ($2.5 \times 10^{10} < M^*/M_{\odot} < 2 \times 10^{11}$;
e.g. \citealt{widrow03}; \citealt{geehan06};
\citealt{hammer07}). Independently of our $D/T$ cut, we find that
approximately $20\%$ of disc galaxies with similar masses as the Milky
Way and M31 have satellite systems with polar alignments relative to
the galaxy.

In Fig. \ref{fig:sats_shape_dmgal2} we show the
probability distribution of the orientation of the satellite galaxies
relative to their host. This differs from the previous calculation, as
we consider all of the satellite galaxies and stack them together
(c.f. \citealt{brainerd05}; \citealt{yang06}). We show the cumulative probability distribution of $\delta$, which is
defined as the angle between the short axis of the host and the
position vector of a satellite galaxy (i.e. $\delta=0$ is polar
alignment and $\delta=90^\circ$ is planar alignment). The blue and red
lines are relative to the minor axis of the galaxy and the minor
axis of the dark matter halo respectively. An isotropic distribution
is shown by the dotted line. The satellites exhibit a roughly planar
alignment relative to the dark matter distribution, in agreement with
the right hand panel of Fig. \ref{fig:sats_shape_dmgal}  where we consider only systems with a large
number of satellites. The satellites have a relatively weaker
alignment relative to the inner disc, although there is a slight bias
towards planar alignment. It is interesting that we find
qualitatively similar results to \cite{brainerd05} and \cite{yang06}
who, owing to the small number of satellite galaxies per parent halo,
use stacked samples of satellites to generate a probability
distribution of their orientations with respect to their
hosts. However, a direct comparison is difficult due to the
different halo selection criteria used by these authors.

\begin{figure*}
  \begin{minipage}{0.33\linewidth}
    \hspace{-0.4cm}
    \includegraphics[width=6.5cm,height=6.5cm]{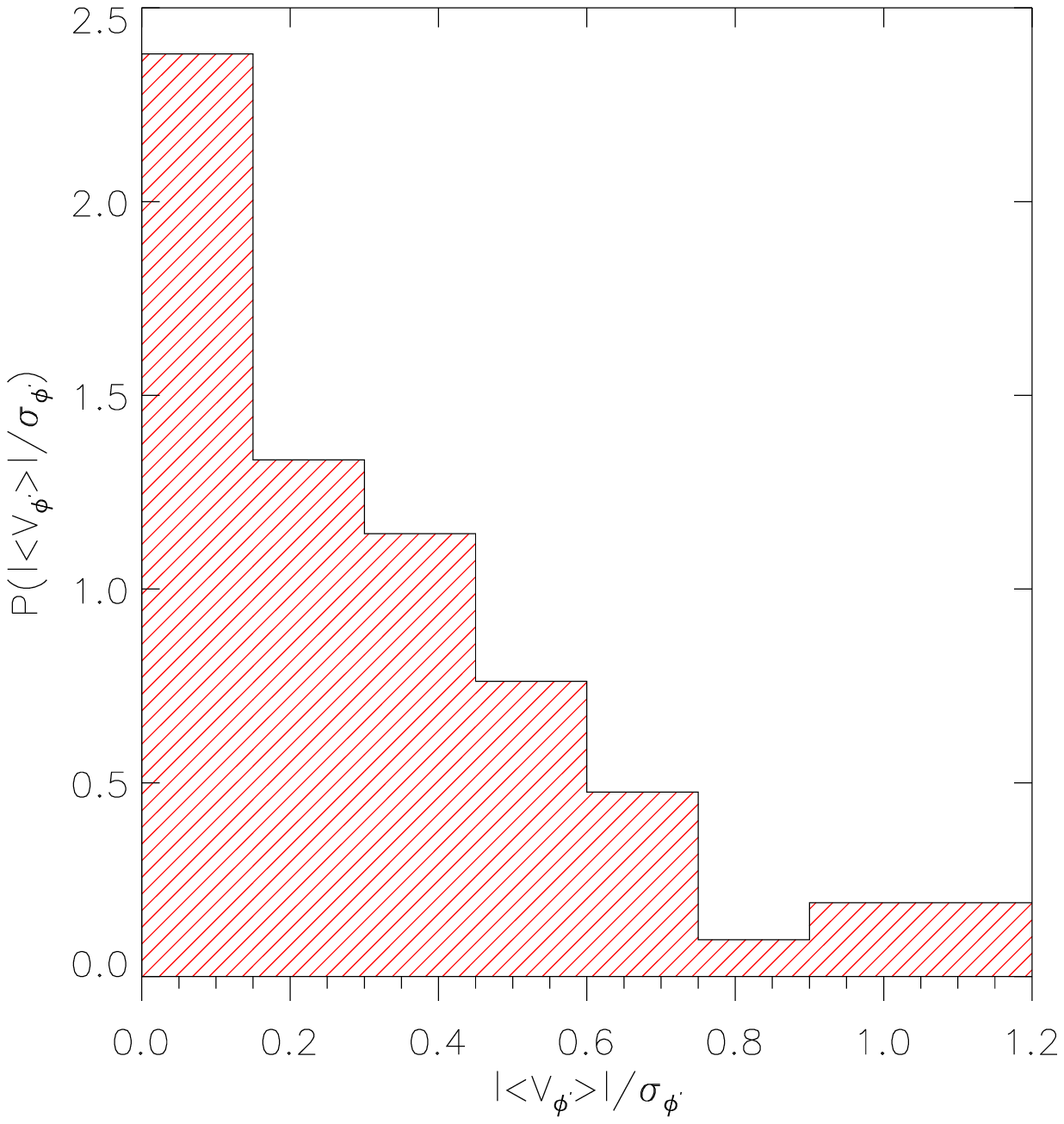}
  \end{minipage}\hfill
  \begin{minipage}{0.33\linewidth}
    \hspace{-0.4cm}
    \includegraphics[width=6.5cm,height=6.5cm]{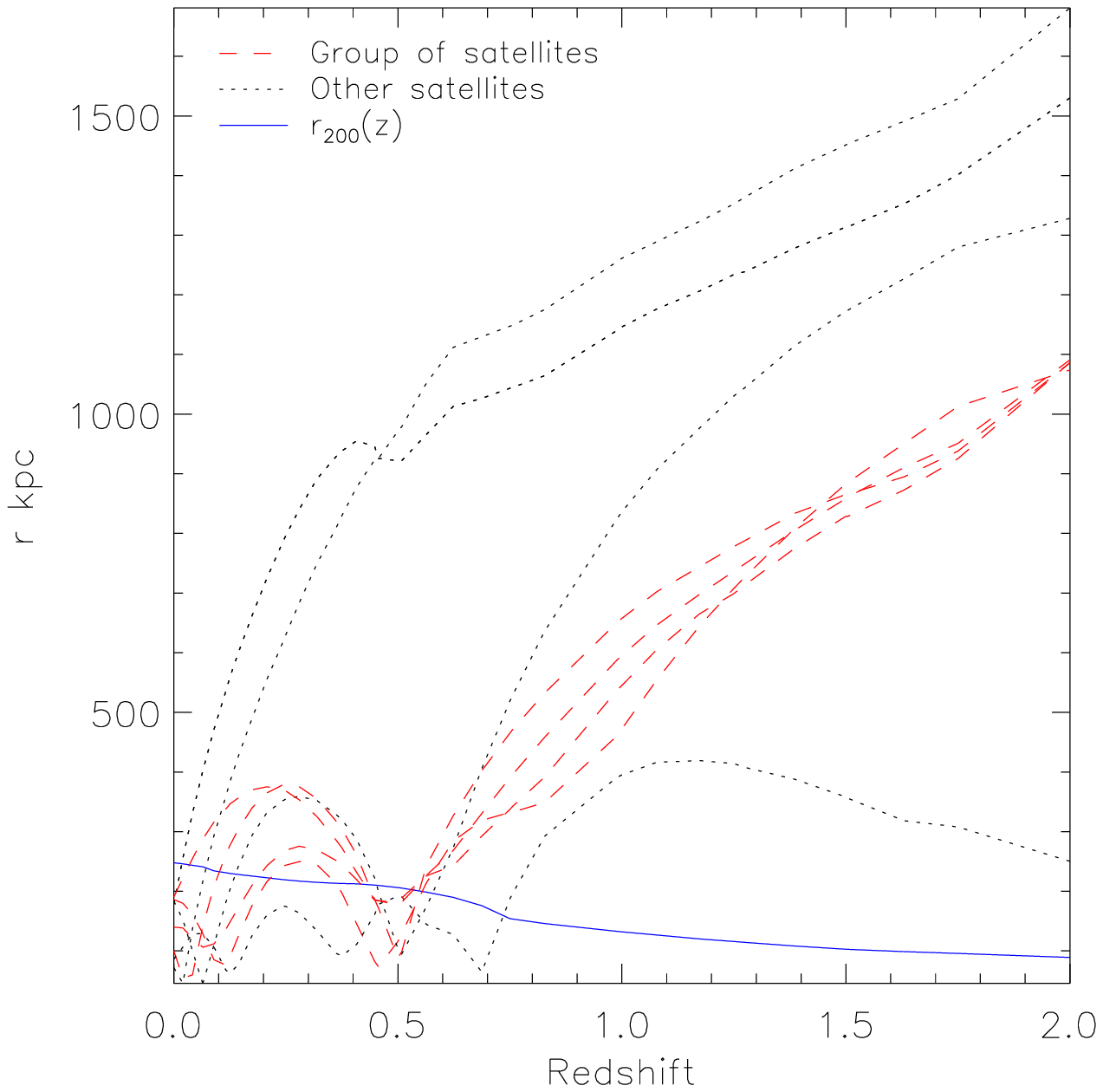}
  \end{minipage}\hfill
   \begin{minipage}{0.33\linewidth}
     \hspace{-0.4cm}
     \includegraphics[width=6.5cm,height=6.5cm]{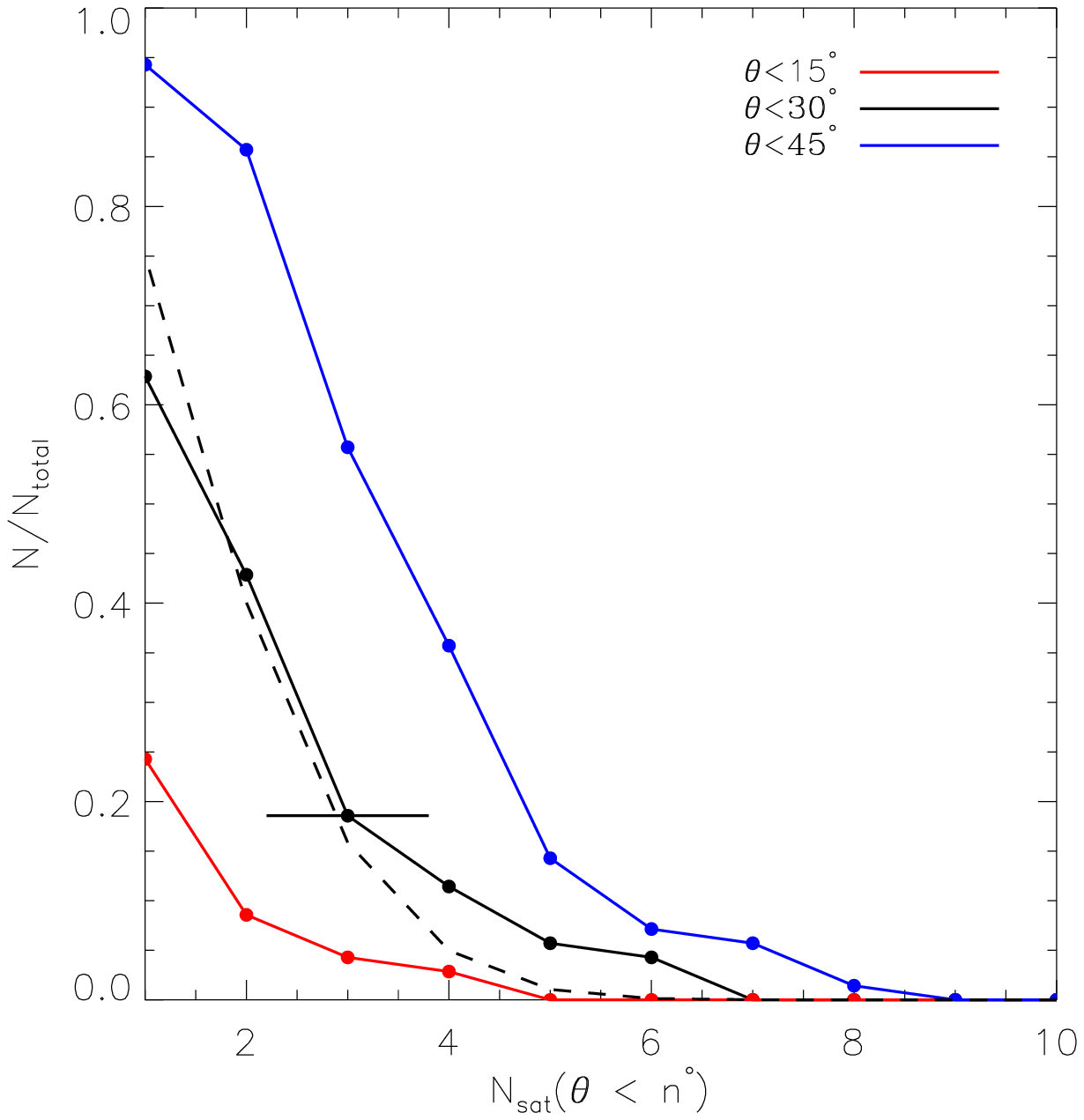}
  \end{minipage}\hfill
  \caption{\small Left-hand panel: The probability distribution,$
      P(|\langle V_{\phi'} \rangle|/\sigma_{\phi'})$, for the ratio
    of rotational versus dispersion support for each system of
    satellites where $N_{\mathrm{sat}}(r \leq r_{200}) > 10$. We only
    consider the ten brightest satellite galaxies within
    $r_{200}$. The median rotation velocity ($\langle
    V_{\phi^{\prime}}\rangle$) and dispersion
    ($\sigma_{\phi^{\prime}}$) are calculated in the plane
    perpendicular to the short axis of the satellite
    distribution. Middle panel: The radial trajectory of a system of
    satellites exhibiting rotational support at $z=0$. A group of
    satellites is accreted at $z \sim 0.6$ (red-dashed lines). This
    group dominates the signature of rotational support shown at
    $z=0$. Dotted lines show the trajectories of some of the other
    satellites and the solid blue line indicates $r_{200}(z)$. Right-hand
    panel: The fraction of systems ($N/N_{\rm total}$) where the
    orbital poles of $N_{\rm sat}$ satellites lie within $15^\circ$
    (red line), $30^\circ$ (black line) and $45^\circ$ (blue line) of
    the short axis defined by the spatial distribution of
    satellites. The horizontal black line indicates the fraction of
    systems which have 3 satellites with orbital poles within $\theta
    < 30^\circ$. The dashed black line shows the case for
      $\theta < 30^\circ$ for an isotropic distribution of satellites
      (in both positional and velocity space).}
  \label{fig:disp_supp}
\end{figure*}

\subsection{Rotational support}
\begin{figure*}
  \centering
  \begin{minipage}{0.9\linewidth}
    \centering
    \includegraphics[width=12cm,height=6cm]{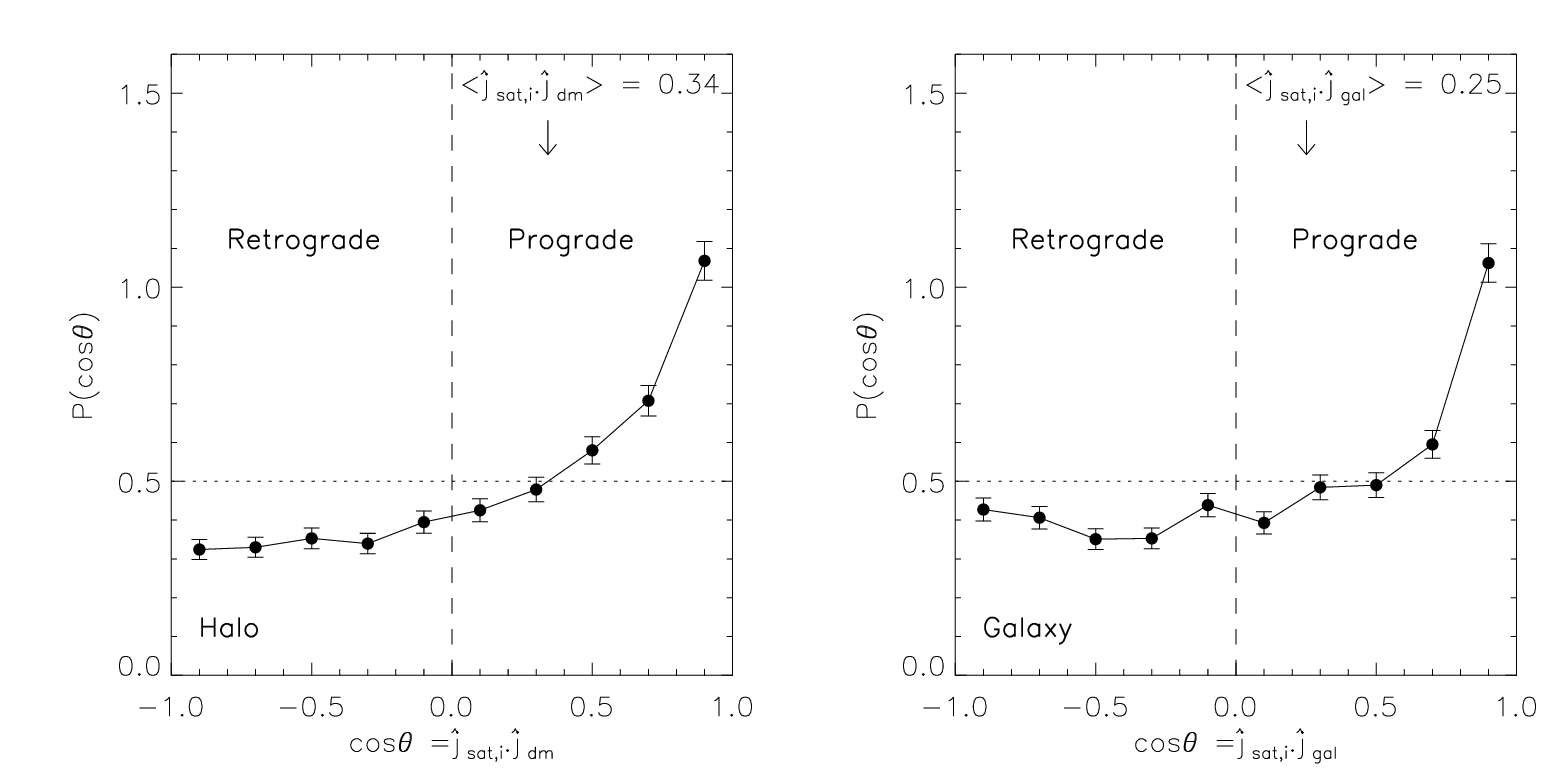}
  \end{minipage}
  \begin{minipage}{0.9\linewidth}
    \centering
    \includegraphics[width=12cm,height=6cm]{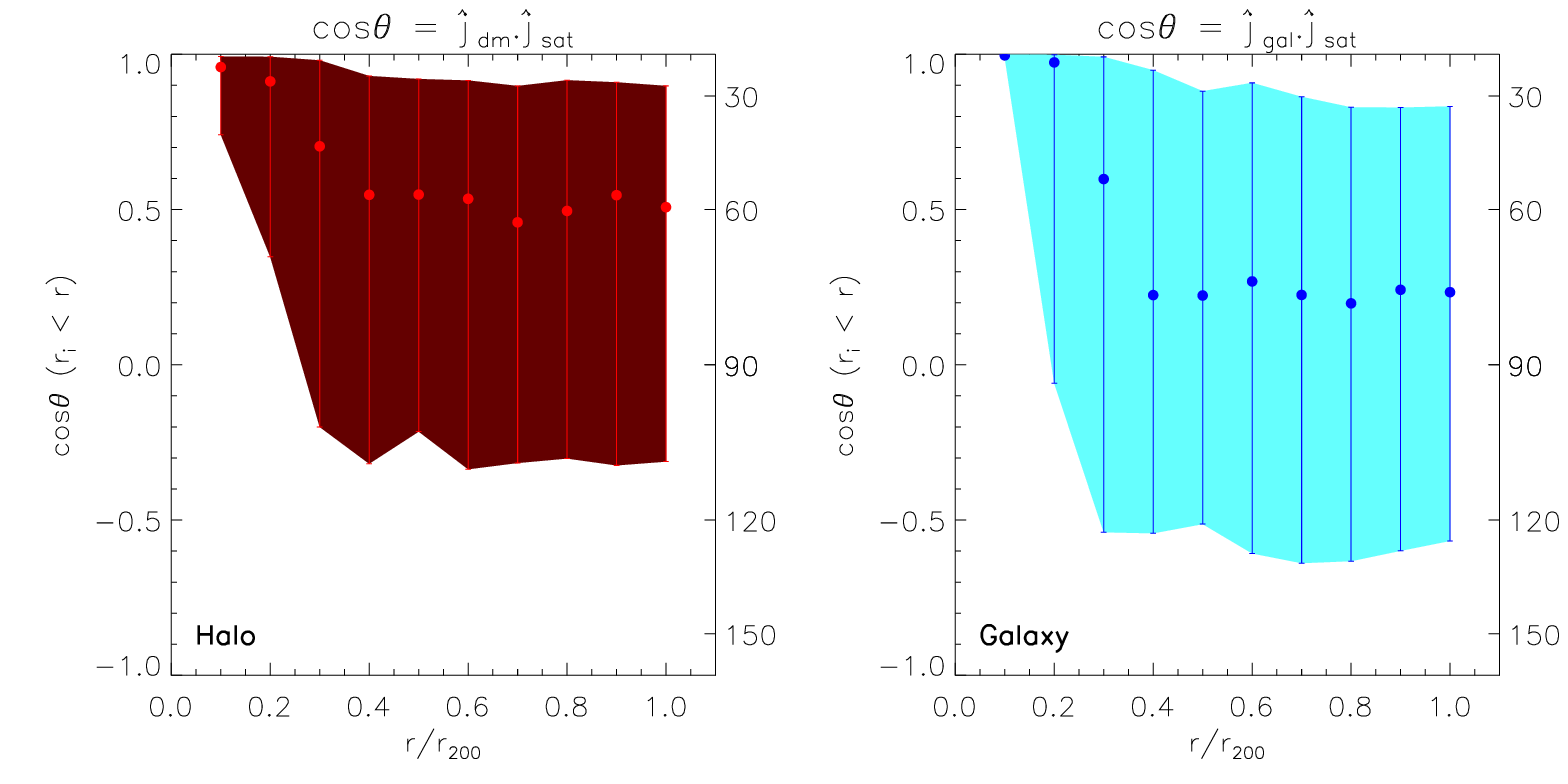}
  \end{minipage}
  \caption{\small Top panels: Alignment between the angular momentum
    vector of each individual satellite galaxy ($\hat{j}_{\rm sat,i}$)
    and its parent
    dark matter halo ($\hat{j}_{\rm dm}(r \le
    r_{200})$, left panel) and galaxy ($\hat{j}_{\rm gal}(r \le
    0.1r_{200})$ right
    panel). Downward pointing arrows give the median values and the
    error bars denote Poisson uncertainties. Bottom panels: The median alignment values between the angular
    momentum vectors of the satellite galaxies and their parent dark
    matter halo (left panel) and and galaxy (right panel) as a
    function of radius. Error bars and shaded regions encompass the
    values covered by 68\% of the distribution. There is a bias
    towards co-rotation (relative to both the spin axis of the dark
    matter halo and the spin axis of the disc) which is more
    pronounced in the inner regions of the halo.}
  \label{fig:satellite_jj}
\end{figure*}

It has been suggested that the satellites of the Milky Way lie in a
rotationally supported disc (e.g. \citealt{kroupa05};
\citealt{metz08}). We calculate the rotational velocity in the plane
perpendicular to the short axis of the satellite distribution
($c_{\mathrm{sat}}$), $\langle v_{\phi^{\prime}} \rangle$ and find the velocity dispersion in this
plane ($\sigma_{\phi^{\prime}}$).  Typical values of these quantities
are:  $\langle v_{\phi^{\prime}} \rangle = 40 \mathrm{kms}^{-1}$ and
$\sigma_{\phi^{\prime}} = 150 \mathrm{kms}^{-1}$. The left-hand panel of Fig. \ref{fig:disp_supp} shows a histogram of the ratio of this
net rotational velocity to the velocity dispersion. Note that here we are restricted to satellite
systems with ten or more members and focus on the ten brightest
satellites within $r_{200}$. The majority of
satellite systems are not rotationally supported and their kinetic
energy is dominated by internal motions. However, there are a small
fraction ($\sim 9\%$) of systems that show substantial rotational
support with $|\langle V_{\phi^{\prime}} \rangle|/\sigma_{\phi^{\prime}} > 0.8$. 

Further investigation into these rotationally supported systems of
satellites suggest that a significant fraction of the satellites are
accreted in groups (i.e. from similar directions at the same time). We
give an example in the middle panel of Fig. \ref{fig:disp_supp}. This
system of satellites defines a highly flattened plane at $z=0$
($s_{\rm sat} \sim 0.25$) and shows evidence of substantial rotational
support with $|\langle V_{\phi^{\prime}}
\rangle|/\sigma_{\phi^{\prime}} \sim 0.9$. The radial trajectories of
some of the satellites belonging to this system are shown in
Fig. \ref{fig:disp_supp}. We can see that 4 of the satellites infall
as a group at $z \sim 0.6$ (or $t_{\mathrm{lookback}} \sim 6
\mathrm{Gyr}$). The coherency of the angular momenta of this group of
satellites is retained until $z=0$ (c.f. \citealt{li08}). Hence,
signatures of rotational support are closely linked to systems of
satellites where a substantial number of the satellite population
today were accreted as a group. \cite{metz09} argue that the
  majority of the Milky Way satellites did not enter the halo in a
  group based on comparisons with dwarf associations in the Local
  Group (found by \citealt{tully06}). Our result suggests a link
  between group infall and rotational support but does not directly
  address the issue of group infall into the Milky Way itself.

\cite{metz08} remark that at least 3 of the classical Milky Way
satellites have orbital poles which lie within $30^\circ$ of the short
axis of the so-called disc of satellites. In the right-hand panel of
Fig. \ref{fig:disp_supp}, we show the distribution of the number of
satellites with orbital poles within $15^\circ$, $30^\circ$ and
$45^\circ$ (red, black and blue lines) of the short axis of their
spatial distribution. In $ \approx 20\%$ of the systems, 3 satellites
have orbital poles within $30^\circ$ of the normal to their spatially
defined distribution. However, no systems have more than 7 (out of 10)
satellites which all have orbital poles within $30^\circ$. Thus, in
agreement with \cite{libeskind09}, we find that an arrangement whereby
the majority of the classical Milky Way satellites have orbital poles
well aligned with the short axis of their flattened distribution will
be difficult to reconcile with the results presented here. For
  comparison, we show with the dashed black line the case for an
  isotropic distribution of satellites in both positional and velocity
  space. The fraction of systems with more than two satellites with
  orbital poles within $30^\circ$ of the short axis of their spatial
  distribution is reduced in the isotropic case. This reinforces the
  point that the satellites are not randomly distributed in phase
  space.

\subsection{Angular Momentum Orientation}
\begin{figure*}
  \centering
  \includegraphics[width=15cm,height=10cm]{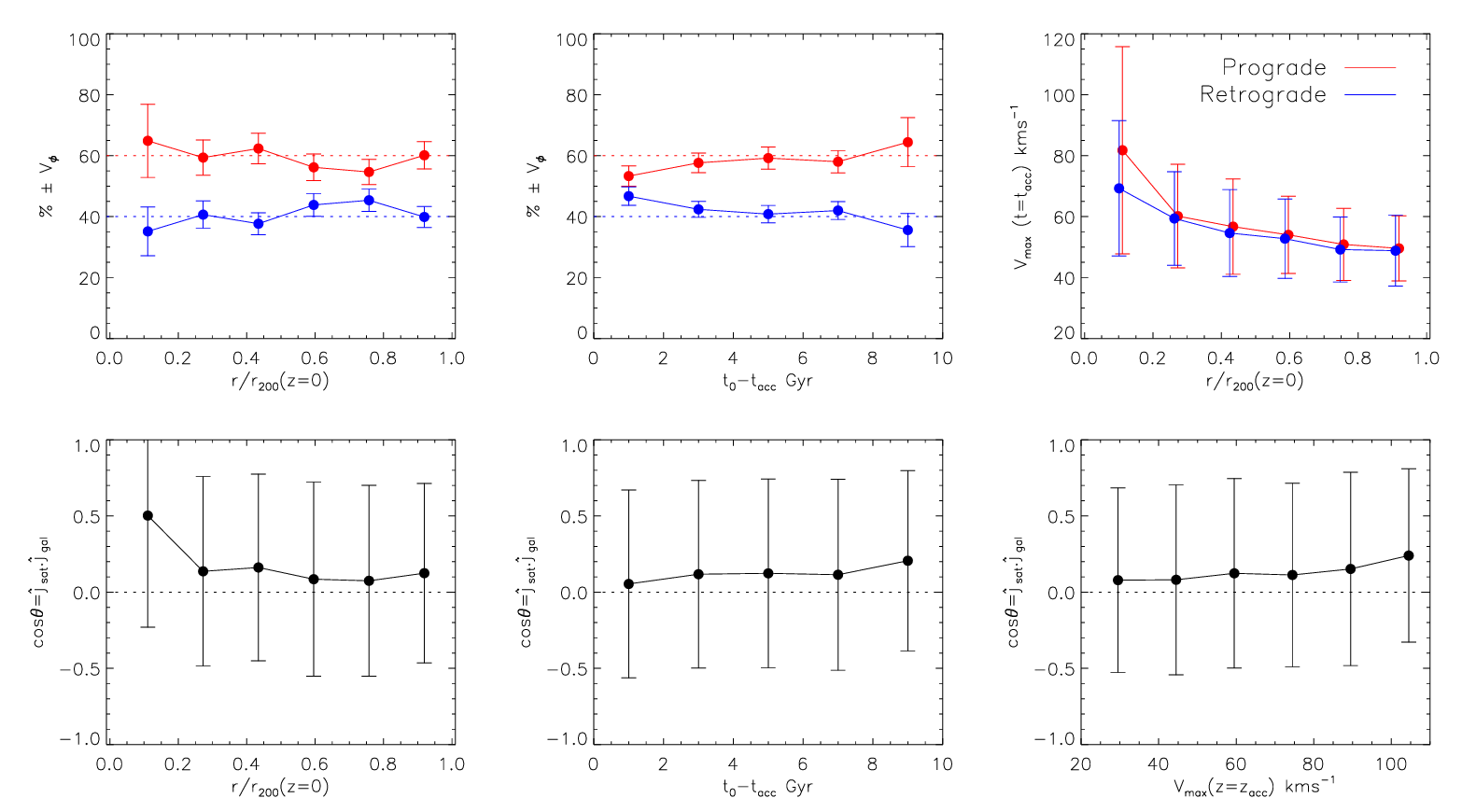}
  \caption{ \small Top left panel: The percentage of satellites on prograde
    (red) and retrograde (blue) orbits as a function of radial
    distance from the parent halo at $z=0$. The dashed horizontal
    lines denote a ratio of 60:40. Error bars give Poisson
    uncertainties. Bottom left panel: The median alignment between the
    satellite angular momenta and the galaxy as a function of
    radial distance at $z=0$. The error bars indicate the dispersion
    in the distributions. Top middle panel: The percentage
    of satellites on prograde (red) and retrograde (blue) orbits as a
    function of accretion time. Bottom middle
    panel: The median alignment between the satellite angular momenta
    and the galaxy as a function of when they were accreted. Top right
    panel: The median satellite $V_{\mathrm{max}}$ at the time of
    accretion as a function of radial distance from the host at
    $z=0$. Bottom right panel: The median alignment
    between the satellite angular momenta and the galaxy as a
    function of satellite $V_{\mathrm{max}}$ at the time of
    accretion. A stronger bias towards prograde orbits is seen by
    satellites accreted at early times and by those satellites which
    were relatively massive at the time of accretion.}
  \label{fig:pro_ret_props}
\end{figure*}

The specific angular momenta of the satellites are computed using
eqn~(\ref{eq:ang_mom}). In this case, we consider our whole sample of
satellite galaxies and are not restricted to parent haloes with a
large number of satellite galaxies.

The top panels of Fig. \ref{fig:satellite_jj} show the distribution of
the angles between the angular momentum vector of each individual
satellite and the angular momentum of the dark matter halo
($\hat{j}_{\rm dm}(r \le r_{200})$, left panel) and the angular
momentum of the galaxy ($\hat{j}_{\rm gal}(r \le 0.1r_{200})$, right
panel). There is an obvious bias in both cases towards alignment
between the angular momentum vectors. With respect to the angular
momentum of the disc (dark matter) approximately 61$\%$ (68$\%$) of
the satellites are on `prograde' orbits ($\mathrm{cos}\theta > 0$) and
39$\%$ (32$\%$) are on `retrograde' orbits ($\mathrm{cos}\theta <
0$). This is in agreement with recent work by \cite{lovell10} who
analysed the orbital angular momentum of dark matter subhaloes in the
Aquarius simulations. All six of the parent haloes in this study
contain populations of co-rotating subhalo orbits. In addition three
of their parent haloes contain subhaloes on retrograde orbits. The
authors attribute these configurations to the filamentary accretion of
subhaloes.

Whilst we see a bias towards co-rotating orbits relative to the inner
galaxy, the spatial orientation of the satellites is only weakly
related to the orientation of the disc. For most satellite systems,
there is little correlation between their shape and net angular
momentum (only a small fraction are rotationally supported). In
addition, phase space mixing is more rapid in spatial coordinates than
velocity space --- the angular momentum orientation of the satellites
will be preserved for longer than any spatial anisotropy. This is
especially true for satellites accreted at earlier times.

In the bottom panels of Fig. \ref{fig:satellite_jj}, we show the
radial dependence of the alignment between the satellite angular
momenta and the dark matter halo angular momenta (left panel) and the
galaxy angular momenta (right panel). The angular momenta of the
satellite galaxies is more closely aligned with the dark matter than
the inner galaxy. The continual accretion of material can alter the
orientation of the net spin of the halo at larger radii. The accretion
of satellites is closely linked to the spinning up of the dark matter
halo (e.g. \citealt{vitvitska02}), hence it is unsurprising that the
satellite angular momenta are well aligned with the dark matter
spin. We found in Section 3.2 that there can be substantial
  misalignments between the angular momentum of the galaxy and the net
  spin of the outer dark matter halo. Hence, the angular momenta of
  the satellite galaxies are less strongly related to the galaxy (this
  is especially true for those satellites accreted at later times).

We find that there is a more pronounced bias towards satellites on
co-rotating orbits (relative to both the disc and dark matter halo) in
the inner regions of the halo. We relate this radial dependence to the
accretion history of the satellite galaxies in the following
section. Note that \cite{hwang10} use a sample of SDSS
  galaxies to study the rotation of satellite galaxies relative to the
  spin of their host galaxy. The authors also find a stronger bias
  towards co-rotating orbits in the inner regions of their host
  galaxies. 

\begin{figure*}
  \centering
   \begin{minipage}{0.45\linewidth}
    \centering
    \includegraphics[width=8cm,height=7cm]{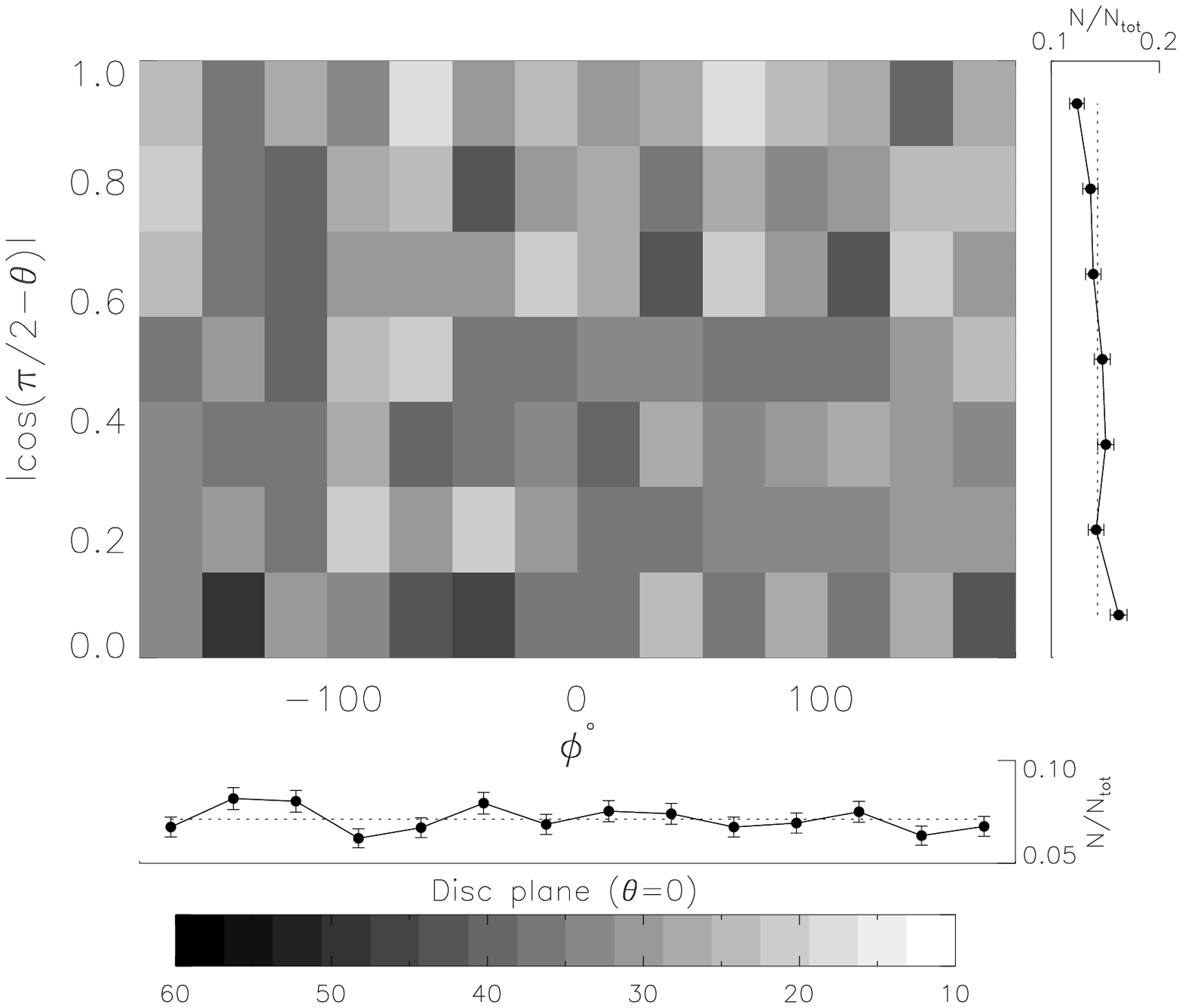}
   \end{minipage}
   \begin{minipage}{0.45\linewidth}
    \centering
    \includegraphics[width=8cm,height=7cm]{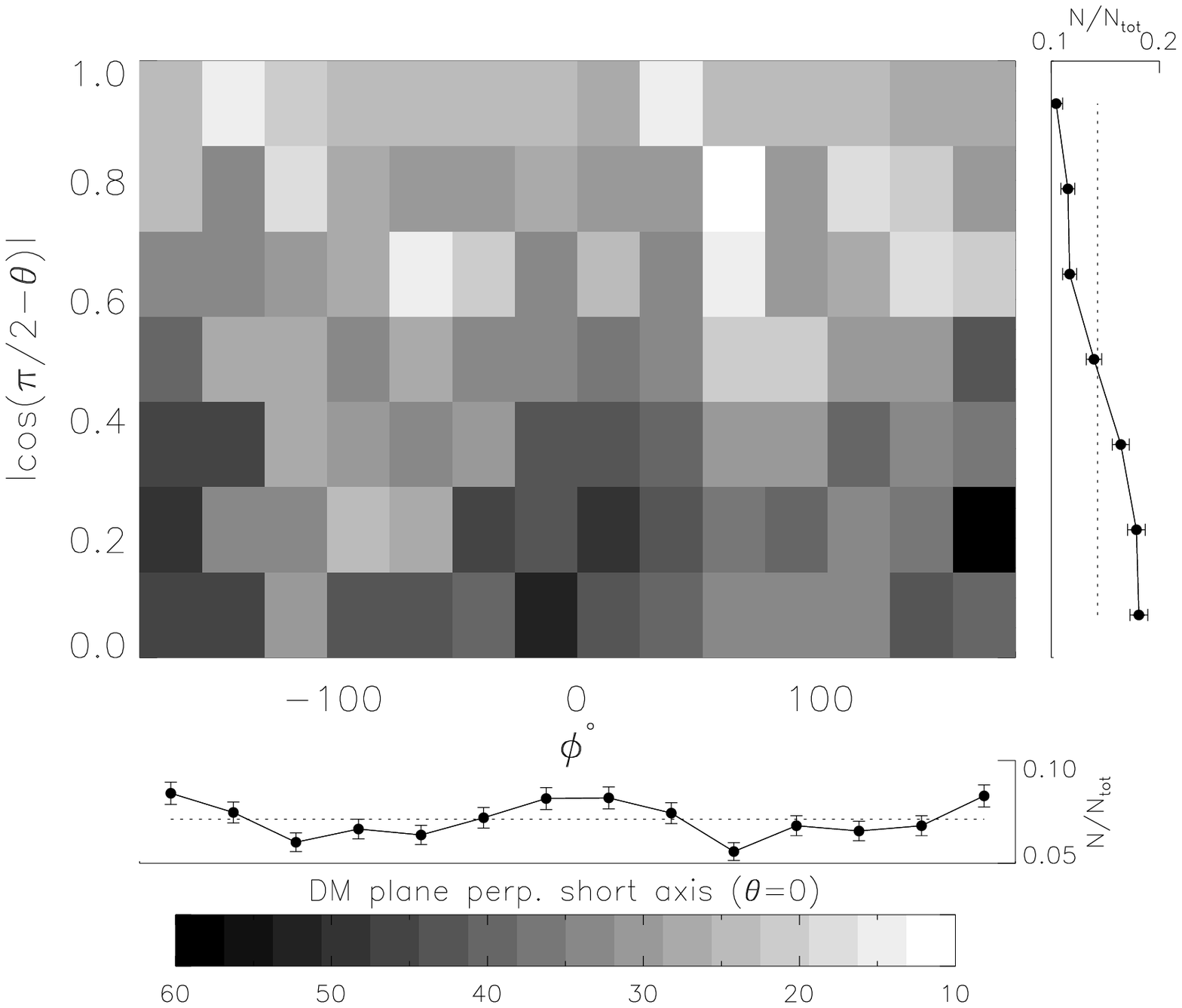}
   \end{minipage}
   \caption{\small The projection in $\mathrm{cos}(\pi/2-\theta)$ and
     $\phi$ of the direction of satellite accretion. $\theta$ and
     $\phi$ are the polar and equatorial axes as defined in the disc
     plane (left panel) at $z=0$ or defined relative to the dark
     matter halo shape (right panel) at $z=0$. The right and bottom
     inset panels give the fractional number of satellites accreted as
     a function of $\theta$ and as a function of $\phi$,
     respectively. Note that the dark matter halo shape is defined for
     $r\le r_{200}$. The positions on the sky are calculated when a
     satellite passes inside $r_{200}(z)$ for the first time. The
     scale bar gives the number of satellites in each pixel. There is
     a preference for satellite accretion in a plane perpendicular to
     the short axis of the dark matter halo (as defined at $z=0$ for
     $r \le r_{200}$) but there is a very weak correlation with the
     galaxy.}
  \label{fig:infall}
\end{figure*}

\begin{figure*}
  \centering
  \includegraphics[width=12cm, height=6cm]{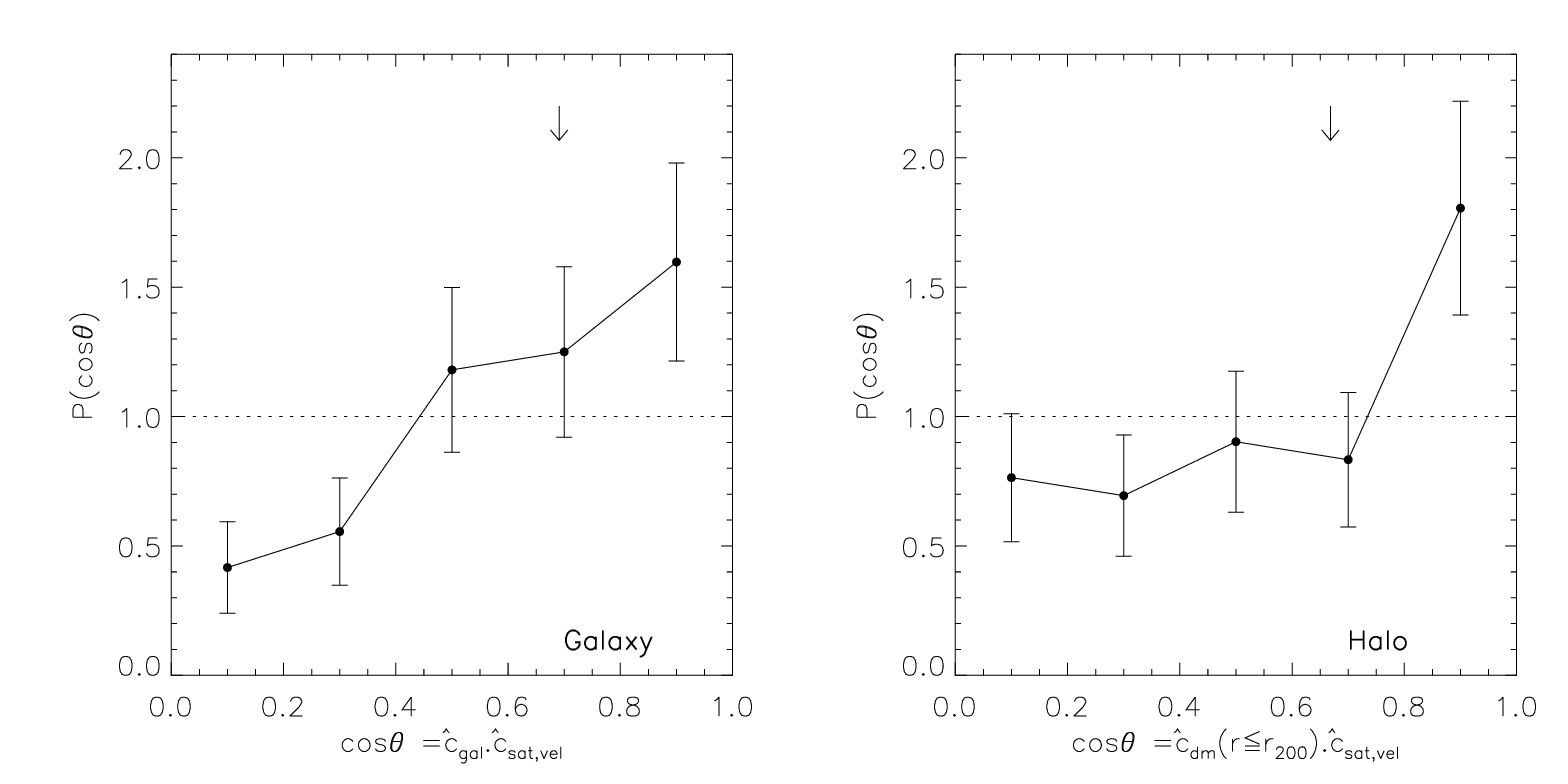}
  \caption{\small The alignment of the axes of the satellite velocity
    anisotropy with the galaxy (left panel) and the dark matter
    halo within $r_{200}$ (right panel). The shape axes of the inner
    galaxy are defined within the region $r \leq 0.1r_{200}$ whilst
    the shape axes of the dark matter distribution are defined for $r
    \leq r_{200}$. The velocity dispersion of the satellites tends to
    be maximum in the plane perpendicular to the short axis of both
    the dark matter halo (right panel) and the galaxy (left
    panel).}
  \label{fig:sat_aniso}
\end{figure*}

\subsection{Anisotropic Accretion}
\label{sec:acc}
The redshift $z=0$ population of satellites is preferentially aligned
in the plane perpendicular to the short axis of the dark matter
distribution and exhibits a bias towards co-rotating orbits (with
respect to both the net spin of the dark matter and the inner
galaxy). Both these observations hint that the satellite galaxies are
accreted from preferential directions (e.g. filaments, \citealt{libeskind05}). According to tidal torque
theory, the angular momentum of a halo is acquired through the tidal
interactions between neighbouring structures (out to $20\mathrm{Mpc}$) and infalling
subhaloes. Presumably some of the surviving population of satellites
today were accreted in a similar fashion and from similar directions
as those subhaloes which originally spun up the halo. Thus, the bias
towards co-rotating orbits can be explained by tidal torque
theory whereby the satellite population today bears the imprint of
those subhaloes which were accreted from similar directions as the
substructures involved in the early stages of galaxy formation.

We trace back our $z=0$ population of satellite galaxies to
$z=2$\footnote{We do not trace our sample of
  satellites beyond redshift $z=2$ as only a very small fraction of
  satellites accreted before this epoch have survived until $z=0$.}. Our
results are summarised in Fig. \ref{fig:pro_ret_props}. Note that we
align our coordinate system with respect to the galaxy at $z=0$
and $\hat{j}_{\mathrm{gal}}$ is defined at $z=0$.  The top left and middle
panels show the percentage of prograde (red) and retrograde (blue)
satellite orbits as a function of the radial distance from the parent
galaxy at $z=0$ and as a function of the time at which the satellite
was accreted. We define the `accretion' time when a satellite first
crosses over $r_{200}(z)$. Similarly, the bottom left and middle
panels show the median misalignment between the galaxy angular
momentum (defined at $z=0$) and the satellites' angular momenta as a
function of radial distance and accretion time. We see a stronger bias
towards prograde orbits for those satellites accreted at larger
lookback times. In addition, those satellites located closer in to the
parent halo have a stronger tendency to be prograde that those located
further out.  The top right hand panel shows the median satellite
$V_{\mathrm{max}}$, the peak of the satellite's circular velocity
profile, at the time of accretion as a function of the
radial distance from the parent galaxy today. Satellites located
closer into the parent halo are on average more massive at the time of
accretion. The bottom right hand panel shows the median alignment
angle between the satellite angular momenta and the galaxy as a
function of the satellite's $V_{\mathrm{max}}$ at the time of
accretion. The most massive satellites at accretion tend to have a
stronger alignment with the net angular momentum of the galaxy.

These results support the suggestion that the bias towards co-rotating
orbits is a consequence of the hierarchical assembly of galaxies. We
find that more massive satellites accreted at earlier times are more
strongly biased towards co-rotation. These satellites resemble the
early substructures which originally spun up the halo. The stronger
bias towards co-rotating satellites in the inner regions of the halo
can be explained by the prevalence of the most massive satellites
\textit{at accretion} in these regions at $z=0$. By dynamical friction
effects, we expect the more massive satellites to sink into the centre
of the parent halo on shorter timescales than those with lower masses.

In Fig. \ref{fig:infall}, we show the 2D projection of the infall
direction of the satellite population at $z=0$. In the left-hand panel,
$\theta$ and $\phi$ are the polar and equatorial axes defined with
respect to the plane of the disc\footnote{The angle $\phi$ is defined
  relative to the major axis of the galaxy but note that as $b/a
  \sim 1$ for a disc this is a rather ambiguous definition}. In the
right-hand panel, $\theta$ and $\phi$ are defined relative to the dark
matter shape (the shape is computed for $r \le r_{200}$). We define a
spherical surface at $r=r_{200}(z)$ and calculate $\theta$ and $\phi$
when a satellite crosses this surface. We stack all the haloes in our
sample where $\theta=0$ defines either the plane of the disc (left
panel), or the plane perpendicular to the short axis of the dark
matter halo (right panel). There is a preference for accretion in a
plane perpendicular to the short axis of the dark matter
distribution. The bottom inset panel shows the fractional number of
satellites accreted as a function of $\phi$ -- this shows there is a
preference for accretion along the major axis of the dark matter halo
(i.e. $|\phi|=0,180^\circ$ or $\pm x$). This is in agreement with
\cite{libeskind05} and \cite{zentner05} who find a preferential direction of satellite
accretion along the major axis of the halo and \cite{bailin05} who
find that there is a strong tendency for the minor axes of haloes to
lie perpendicular to large-scale filaments. There is a much weaker
bias relative to the disc plane. In Section \ref{sec:sat_aniso}, we
found that the spatial orientation of the satellite galaxies at $z=0$
is much more strongly correlated with the dark matter shape rather
than the galaxy. Note that both the disc plane and short axis of
the dark matter halo are defined at $z=0$. We note that this does not
take into account the evolution of the orientation of the galaxy
or dark matter halo. However, our analysis does show that when
satellites are accreted they have a preferential alignment relative to
the dark matter halo \textit{as defined today}.

\subsection{Velocity Anisotropy}

\begin{figure}
  \centering
  \includegraphics[width=8.5cm,height=8.5cm]{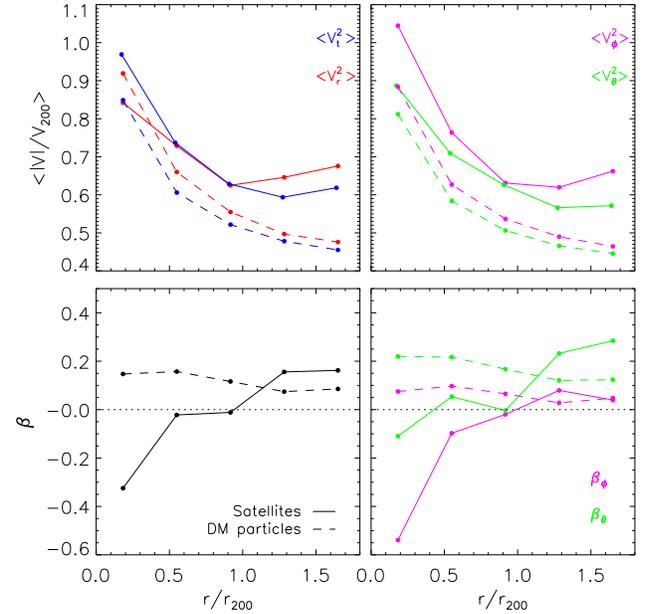}
  \caption{\small Top-left panel: The root-mean-square radial
    ($V_r$, red) and
    tangential ($V_t$, blue) velocity components as a function of radius. The
    solid and dashed lines are for the satellite galaxies and dark
    matter particles respectively. Top-right panel: The root-mean-square equatorial ($V_\phi$, magenta) and
    polar ($V_\theta$, green) components as a function of radius. Bottom-left panel: The velocity
    anisotropy parameter ($\beta$) as a function of radius. $\beta$ is
    consistent with isotropy over a wide radial range
    Bottom-right panel: The polar and equatorial velocity anisotropy
    components ($\beta_\theta$, $\beta_\phi$) as a function of
    radius.}
  \label{fig:beta}
\end{figure}

\begin{figure*}
  \centering
  \includegraphics[width=12cm,height=6cm]{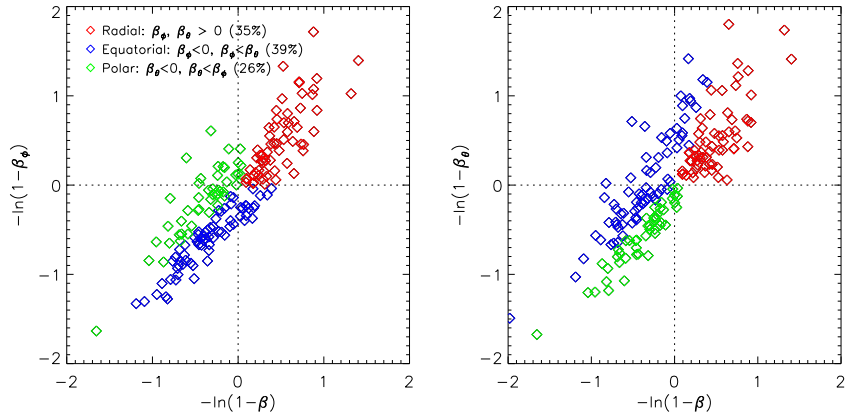}
  \caption{\small The polar ($\beta_\theta$, right panel) and equatorial
    ($\beta_\phi$,left panel) velocity anisotropy parameters against
    the overall anisotropy parameter, $\beta$. Each point represents a
    system of ten or more satellites (within $2r_{200}$) belonging to
    an individual galaxy. Red, blue and green points represent orbits
    dominated by radial, equatorial and polar motions
    respectively. There is significant halo-to-halo scatter in the
    overall velocity anisotropy.}
  \label{fig:beta_pt}
\end{figure*}
By restricting attention to those systems which have ten or more
satellite galaxies, we now examine the properties of the velocity
dispersion tensor.  The velocity anisotropy of actual satellite galaxy
populations has never been measured, but it is an important parameter
for studies which use satellites as tracers of the dark matter
potential. Fig.\ref{fig:sat_aniso} shows the alignment of the
principal axes of the velocity dispersion tensor with the shape axes
of the galaxy and the dark matter halo.

The short axis of the velocity anisotropy tensor tends to align with
the short axis of the galaxy (and the short axis of the dark
matter distribution). This suggests the velocity dispersion of the
satellites tends to be maximum in the plane perpendicular to the short
axis, i.e. in the plane of the disc. This cylindrical alignment of the
velocity anisotropy tensor agrees with our earlier findings that there
is a bias toward co-rotating satellite orbits.  Note that here we have
no direct comparison to the Milky Way satellites as their tangential
velocity components are poorly constrained.

The anisotropy parameter, $\beta$ is defined as
\begin{equation}
\beta=1-\frac{\langle V^2_t \rangle}{2\langle V^2_r \rangle}=1-\frac{\langle V^2_\theta \rangle+\langle V^2_\phi \rangle}{2\langle V^2_r \rangle}.
\end{equation}
This is a measure of how radially or tangentially biased the satellite
orbits are. To distinguish between polar and equatorial biased orbits,
we define two more anisotropy parameters
\begin{equation}
\beta_\theta=1-\frac{\langle V^2_\theta \rangle}{\langle V^2_R
  \rangle}, \; \; \;
\beta_\phi=1-\frac{\langle V^2_\phi \rangle}{\langle V^2_R
  \rangle}
\end{equation}
These quantities are defined in a cylindrical polar coordinate system
($R,\phi,z$) aligned such that the $z$-axis is normal to the disc.

In the top-left panel of Fig. \ref{fig:beta}, we show the radial
dependence of the radial and tangential velocity components. The solid
lines are for the satellite galaxies, whilst the dashed lines are for
the dark matter. Radial velocities dominate over tangential
velocities at all radii for the dark matter particles. The satellite
galaxies have more tangentially biased orbits in the inner regions
($r<0.5r_{200}$) and only become radially biased at larger radii
($r>r_{200}$). Over a wide radial range, the satellites are consistent
with an isotropic velocity distribution. This can also be seen in the
radial dependence of $\beta$ shown in the bottom-left panel. The two
right-hand panels decompose the tangential velocity component into its
polar and equatorial parts. The tangential bias of the satellite
orbits in the inner regions are dominated by their equatorial
motion. This is seen in the bottom-right panel where $\beta_\phi$ is
(comparatively) large and negative at small radii. The equatorial
component dominates over the polar component for the dark matter
particles over the whole radial range. Dark matter haloes are
dispersion supported (see Fig. \ref{fig:halo_align}), and the
dominance of equatorial velocity components over the polar components
leads to the (slight) flattening of the dark matter halo in the
z-direction\footnote{Note the flattening of the dark matter halo is
  not always in the z-direction (defined relative to the inner disc)
  as there can be misalignments between the short axis of the dark
  matter distribution and the z-axis of the galaxy, especially
  at larger radii (see Section \ref{sec:halo_props}). However, there
  is reasonably good alignment for the majority of haloes.}.

In Fig. \ref{fig:beta}, we have stacked all the satellites in our
sample together by normalising radial distances by $r_{200}$ and
velocities by $V_{200}$. For parent haloes with ten or more
satellites within $2r_{200}$, we plot the velocity anisotropy parameters, $\beta$,
$\beta_\theta$ and $\beta_\phi$ in Fig. \ref{fig:beta_pt}. It is
apparent that there is a wide spread in the $\beta$ parameters for
individual haloes ranging from $-2 < \beta < 0.8$. By stacking all the
satellites of different haloes together, we find $\beta$ values ranging
from $-0.4 < \beta < 0.2$ over the same radial range. We can see from
Fig. \ref{fig:beta_pt} that there is a slight bias towards satellite
systems dominated by motion in the equatorial plane ($39\%$) but many satellite
systems are dominated by their radial motions ($35\%$) or are biased towards
polar orbits ($26\%$).

The velocity anisotropy parameter, $\beta$, is an important factor
required to estimate the masses of local group galaxies such as the Milky
Way and M31 (see next Section \ref{sec:mest}). This parameter is
difficult to observe as in most cases only line of sight velocities
are available and we lack full 3D velocity information. To overcome
this, many authors adopt anisotropy parameters derived from
simulations (e.g. \citealt{xue08}; \citealt{watkins10}). Our finding
that there is significant halo-to-halo scatter means that significant
caution is warranted when applying a velocity anisotropy applicable to
a simulated satellite system of an individual halo \textit{or} from a
stacked sample of satellites to our own Milky Way Galaxy.

\begin{figure}
  \centering
  \includegraphics[height=8cm]{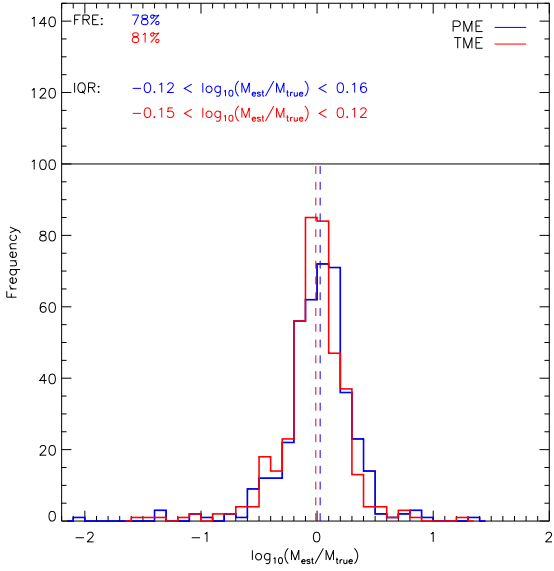}
  \caption{\small Histograms showing the ratio of the estimated mass
    to the true mass for the two different mass estimators. The
    projected mass estimator (PME) and Tracer mass estimator (TME) are
    shown by the blue and red lines and respectively. We give the fraction
    of reasonable estimates (FRE) and the inter-quartile range (IQR)
    as defined in the text.}
  \label{fig:iqr}
\end{figure}

\section{An application: Mass Estimators}
\label{sec:mest}

In practice, simple estimators are often used to compute the mass of a
dark halo from the positions and velocities of the satellite galaxies
(e.g., Watkins et al. 2010).  They depend on simplifying assumptions,
such as underlying spherical symmetry of the dark halo, or constant
velocity anisotropy. As we have seen, there are numerous effects
present in the simulations -- triaxiality of the halo, continuing
infall to the present day, variation of anisotropy with radius -- that
are not accounted for in the mass estimators.  Hence, it is
interesting to see how the estimators fare against simulation data.

The projected mass estimator (hereafter PME) introduced by
\cite{bahcall81} takes the form:
\begin{equation}
  M=
  \frac{C_p}{GN}\sum_{i=1}^{N}v_{\mathrm{los},i}^2R_i,
\label{eq:pme}
\end{equation}
where $R$ and $v_{\mathrm{los}}$ are the projected positions and line-of-sight velocities of the $N$ satellite galaxies.  The constant is
$C_p=16/\pi$ (isotropic) or $C_p=32/\pi$ (radial orbits).

The tracer mass estimator (hereafter TME) is given in Watkins et
al. (2010, see also Evans et al. 2003). It assumes a spherically
symmetric power-law for the halo potential $\Phi \propto
r^{-\alpha}$. We use the form:
\begin{equation}
M=\frac{C_\mathrm{T}}{GN}\sum_{i=1}^{N}v_{\mathrm{los},i}^2R_i^{\alpha}
\label{eq:tme}
\end{equation}
where $\alpha \sim 0.5$ and the constant $C_\mathrm{T}$ is given in
eqn (26) of Watkins et al. (2010) and depends on the velocity
anisotropy $\beta$. In particular, $\Phi \propto r^{-0.5}$ is a good
approximation to a Navarro-Frenk-White (NFW, \citealt{navarro96}; \citealt{navarro97}) profile often used to
describe simulated dark matter haloes. We assume isotropic orbits
($\beta=0$), but investigate the validity of this assumption by
comparing the estimated masses when the true velocity anisotropy of
the tracer satellites is used.

Eqns~(\ref{eq:pme}) and (\ref{eq:tme}) provide an estimate for the
total mass within the radius of the furthest tracer
($r_{\mathrm{out}}$).  We select an arbitrary viewing angle for our
simulated haloes to generate projected positions and line of sight
velocities.  We then compute the `true' mass within $r_{\mathrm{out}}$
for each halo and compare to masses found via the two mass
estimators. Note we only select satellites within $r < 1.5r_{200}$. We
use all satellites, but check that our results are not significantly
affected when only luminous satellites are included

In Fig. \ref{fig:iqr}, we present histograms of the ratio between the
estimated and true mass. We define the `Fraction of Reasonable
Estimates' or FRE as the fraction of estimates in the range $0.5 <
M_{\mathrm{est}}/M_{\mathrm{true}} < 2$. We also give the Inter
Quartile Range (IQR) of the mass estimates in Figure
\ref{fig:iqr}. Both the TME and the PME perform well, with $\approx
81\%$ and $78\%$ of the estimates satisfying our `reasonable'
criteria, respectively.\footnote{By restricting ourselves to systems
  with 10 or more satellites the fraction of reasonable estimates
  increases to $\sim 90\%$.}. The TME performs slightly better than
the PME (note the more symmetrical distribution in
Fig. \ref{fig:iqr}), but both provide good results especially
considering the rather idealized assumptions under which they are
derived. The IQR shows that the uncertainty in the mass estimates
given by the simulations is of similar magnitude ($\sim 30\%$) to the statistical
uncertainty found by \cite{watkins10} in their estimates of the Milky
Way mass.

\begin{figure*}
  \centering
  \includegraphics[width=12cm, height=6cm]{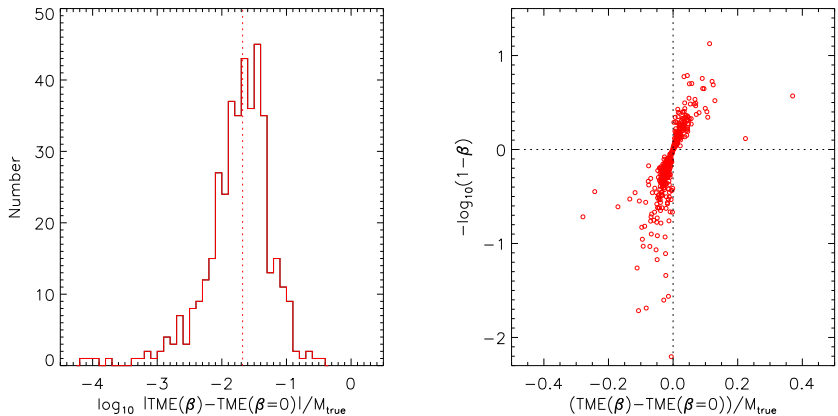}
  \caption{\small Left hand panel: Histogram showing the absolute
    difference in the estimated mass when the actual velocity
    anisotropy is used instead of assuming $\beta=0$. This is
    normalised by the true mass. Right hand panel: The variation in
    the mass estimator when the true velocity anisotropy is used as a
    function of velocity anisotropy.}
  \label{fig:mest_beta}
\end{figure*}

Fig. \ref{fig:mest_beta} illustrates how the mass given by the TME
varies when we use the true velocity anisotropy parameter instead of
assuming an isotropic distribution. Assuming isotropy for tangential
orbits leads to an overestimate in the mass, whilst the reverse is
true for radial orbits. The median absolute difference between the
isotropic estimator and the true anisotropy estimator is $2\%
M_{\mathrm{true}}$. Unless the velocity anisotropy is strongly radial
or tangential, the assumption of isotropy yields reasonable mass
estimates. This is important for the application of such estimators to
observational data, as we rarely have observationally derived values
of the velocity anisotropy parameter.

\begin{figure*}
  \centering
  \includegraphics[width=12cm, height=6cm]{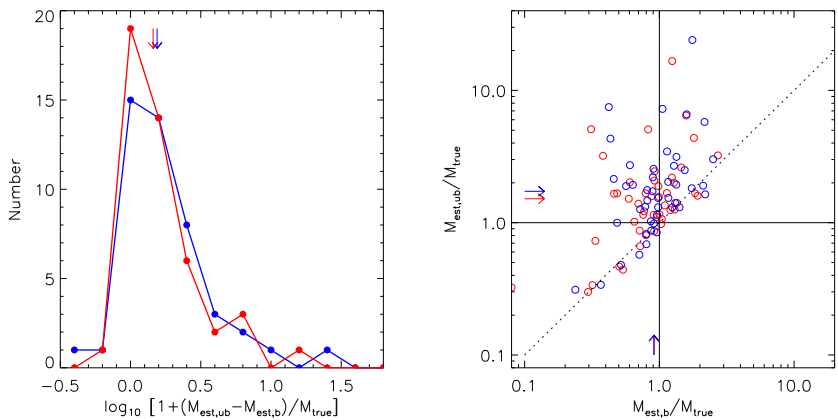}
  \caption{\small Left hand panel: Number distribution showing the
    relative difference in the mass estimates when unbound satellites
    are included or excluded in the analysis. Median values are shown
    by downward pointing arrows. Right hand panel: The
    ratio between the estimated mass and true mass in the case where
    unbound satellites are included (y-axis) and when they are not
    (x-axis). The colour scheme is the same as Fig. \ref{fig:iqr}. Arrows
    represent the median values.}
  \label{fig:mest_unbound}
\end{figure*}

In our samples of satellite galaxies, approximately 3$\%$ are
unbound. Whilst unbound satellites are not common, their inclusion in
mass estimators can cause fairly large deviations from the true
mass value.  We compute the estimated mass with and without the
unbound satellites for the 46 haloes which contain at least one
unbound satellite (within $1.5r_{200}$).  The left hand panel of
Fig.~\ref{fig:mest_unbound} shows the distribution of the relative
difference between the mass estimates. The right hand panel plots the
two mass estimates against one another. Inclusion of unbound
satellites causes a systematic overestimate of the mass. The TME and
PME have median differences of $45\% M_{\mathrm{true}}$ and $55\%
M_{\mathrm{true}}$ when unbound satellites are included.

\section{Conclusions}

We investigated the orbital properties of the satellites of late-type
galaxies using the \textsc{gimic} suite of simulations
(\citealt{crain09}). These state-of-the-art simulations incorporate
baryonic physics into a $\Lambda$CDM cosmological framework and
produce realistic disc galaxies at $z=0$. We analyse the phase space
distributions of the satellite galaxies relative to the luminous
baryonic material (i.e. the central galaxy disc) as well as the unseen
dark matter component with a large sample of galaxies. In this way we
can provide a more direct comparison with observations. Our sample of
parent haloes were chosen to be relaxed systems in the mass range $5
\times 10^{11} < M_{200}/M_{\odot} < 5 \times 10^{12}$, which broadly
overlaps with the mass of our own Milky Way galaxy and of M31.

The parent dark matter haloes in our sample are generally triaxial but
roughly spherical and have axial ratios which are roughly constant
with radius ($ \langle s_{\rm dm} \rangle \sim 0.8$). Comparison to
the dark matter only counterparts of our sample shows that the
inclusion of baryonic physics affects the shapes of the dark matter
haloes significantly, even out to $r_{200}$.  The central galaxy (or
`disc') is often misaligned in both shape and angular momentum with
the dark matter halo. The inner regions of the dark matter halo are
well aligned with the central galaxy but there can be substantial
misalignments at larger radii, in the region most relevant for the
satellite galaxies. We find that for radii $r \sim r_{200}$, the short
axis of approximately 30\% of our parent dark matter haloes are
significantly misaligned from the short axis (or z-direction) of the
inner galaxy ($\theta > 45^\circ$). In velocity space, the net spin of
the dark matter halo can be almost perpendicular to the angular
momentum vector of the inner galaxy in $\sim 40\%$ of our sample and
$2\%$ are even spinning in the opposite sense to the inner galaxy.

There is an obvious spatial bias between the dark matter of the parent
halo and the satellite galaxies. The satellite system has a more
flattened shape than the dark matter halo and is not as centrally
concentrated.  By considering all systems with 10 or more satellites
within $r_{200}$, we find that the satellites preferentially align in
a plane perpendicular to the short axis of the dark matter
halo. However, owing to the misalignments between the inner galaxy and
(outer) dark matter halo, this preferential alignment is much weaker
relative to the inner galaxy (cf. \citealt{agustsson06}). In fact, by
only considering the 10 highest stellar mass satellites in each
system, we find the distribution of satellites is almost uniform
relative to the central disc and the `unusual' orientation of the
classical Milky Way dwarfs is not uncommon: 20$\%$ of satellite
systems are perpendicular (within $10^\circ$) to the disc. In a
similar fashion to \cite{brainerd05}, we find the probability
distribution of the orientation of the satellite galaxies relative to
their hosts by stacking \textit{all} of the satellites in our
sample. In qualitative agreement with the observational results from
SDSS and the 2dF Galaxy Redshift Survey, we find that the satellites
have a weak bias towards planar alignment relative to the
disc. However, there is a much stronger alignment relative to the dark
matter halo shape.

It has been suggested that the Milky Way satellites may occupy a
rotationally supported disc (\citealt{metz08}). We find that satellite
systems which are planar and rotationally supported are relatively
uncommon ($\sim 9\%$) in the simulations. This often occurs when a
large fraction of the satellites is accreted in a group that retains
its coherence in velocity space until $z=0$. We find that it is not
unusual to have 3 out of 10 satellites with orbital poles within
$30^{\circ}$ of the normal to their spatial configuration. This is
consistent with the available proper motion data on the classical
dwarfs. However, we find that if a substantial number of the classical
dwarfs of the Milky Way (e.g. 7 out of 10) were found to have orbital
poles aligned with the normal to the disc of satellites, then this
would be inconsistent with the results of our simulations.

We tested this claim in the simulations and found that satellite
systems which are planar and rotationally supported are relatively
uncommon ($\sim 9\%$). This often occurs when a large fraction of the
satellites is accreted in a group that retains its coherence in
velocity space until $z=0$. We find that it is not unusual to have 3
out of 10 satellites with orbital poles within $30^{\circ}$ of the
normal to their spatial configuration. This is consistent with the
available proper motion data on the classical dwarfs. However, we find
that if a substantial number of the classical dwarfs of the Milky Way
(e.g. 7 out of 10) were found to have orbital poles aligned with the
normal to the disc of satellites, then this would be inconsistent with
the results of our simulations.

There is a bias towards co-rotating satellite orbits relative to both
the angular momentum of the disc and the net spin of the dark matter
halo. This is more pronounced in the inner regions of the halo. This
confirms earlier results relating to dark matter only simulations
(e.g. \citealt{lovell10}), but the bias with respect to the inner disc
is weaker owing to the angular momentum misalignments between the
inner galaxy and dark matter halo. A preference for co-rotating orbits
is a natural consequence of the hierarchical assembly of galaxies
whereby satellites accreted at earlier times are related to those
substructures that helped spin up the galaxy. We confirmed this by
finding a stronger bias towards prograde orbits by the more massive
satellites that were accreted at earlier times. By tracing back the
infall orientation of our sample of satellite galaxies, we find their
anisotropic distribution is due to their preferential accretion in
directions perpendicular to the short axis of the dark matter
distribution. There is a weaker correlation with the orientation of
the inner galaxy.

The velocity anisotropy tensor for the satellite galaxy systems is
cylindrically aligned relative to the central disc. Tangential motions
dominate at smaller radii, often due to the prevalence of equatorial
(as opposed to polar) orbits. The velocity anisotropy, $\beta$, is an
important parameter which is largely inaccessible with present
observations.  Here we show that $\beta$ is consistent with zero over
a large radial range when all satellites are stacked
together. However, inspection on a halo-by-halo basis shows that there
is a significant degree of scatter between haloes. This scatter puts
into question the validity of using a single simulation as a template
velocity anisotropy to be applied to real galaxies.

Finally, we considered an application of the orbital properties of the
satellite galaxies. We tested two popular mass estimators in the
literature which make use of the projected positions and line of sight
velocities of tracers, such as satellite galaxies, to estimate the
mass of the parent halo. The projected mass estimator (PME) and the
tracer mass estimator (TME) both perform well and estimate
`reasonable' (within a factor of 2 of the true mass) halo masses. The
TME performs slightly better as it assumes the satellites are tracers
of the halo potential rather than having their density profile
generated by the dark matter potential (as assumed by the PME). We
found that an unknown velocity anisotropy parameter can lead to
incorrect mass estimates but these are only substantial when $\beta$
is significantly non-isotropic. In addition, the inclusion of unbound
satellites can cause large overestimates of the true halo mass.
\\
\\
\noindent
Requests for simulation data should be directed to
ajd75@ast.cam.ac.uk.

\section*{Acknowledgments}
AJD thanks the Science and Technology Facilities Council (STFC) for
the award of a studentship, whilst VB and AF acknowledge financial
support from the Royal Society. IGM is supported by a Kavli Institute
Fellowship at the University of Cambridge.  CSF acknowledges a Royal
Society Wolfson Research Merit award. RAC is supported by the
Australian Research Council via a Discovery Project grant. The
simulations presented here were carried out using the HPCx facility at
the Edinburgh Parallel Computing Centre (EPCC) as part of the EC's
DEISA 'Extreme Computing Initiative', the Cosmology Machine at the
Institute for Computational Cosmology of Durham University, and on the
HPC Cluster Darwin at the University of Cambridge. We wish to thank
Joop Schaye and Volker Springel for help and advice.

\bibliography{mybib}

\begin{appendix}

\section{Convergence tests}

\begin{figure*}
  \centering
  \includegraphics[width=12cm,height=6cm]{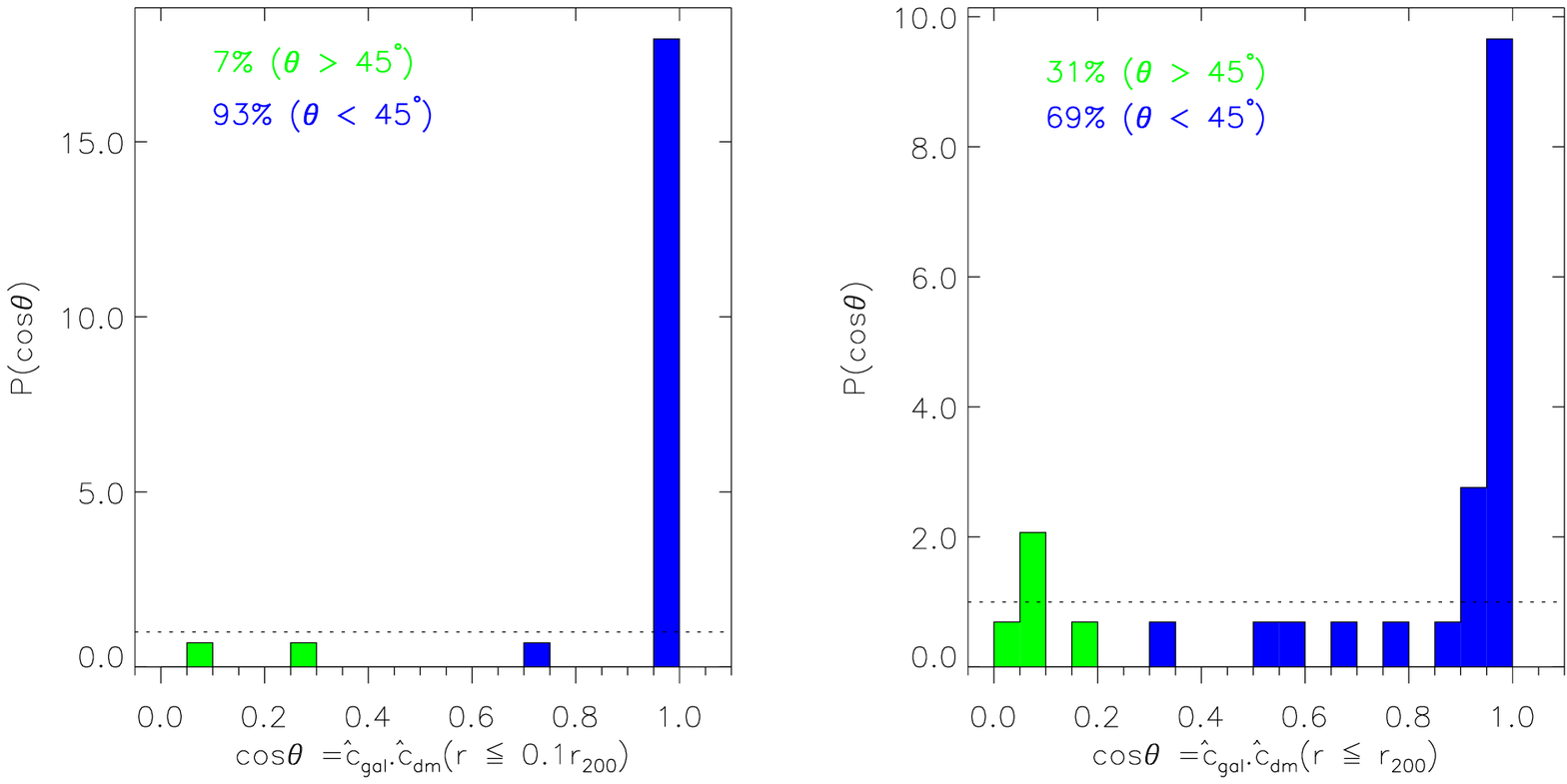}
  
   \caption{\small  The distribution
    of misalignment angles between the short axis of the galaxy
    and the short axis of the dark matter halo for $r \le 0.1r_{200}$
    (left panel) and $r \le r_{200}$ (right panel) respectively. This
    is for the high resolution \textsc{gimic} simulations
    (c.f. Fig. \ref{fig:histogram} for the intermediate resolution version
    of this plot).}
    \label{fig:app1}
\end{figure*}

\begin{figure*}
  \centering
  \includegraphics[width=12cm,height=6cm]{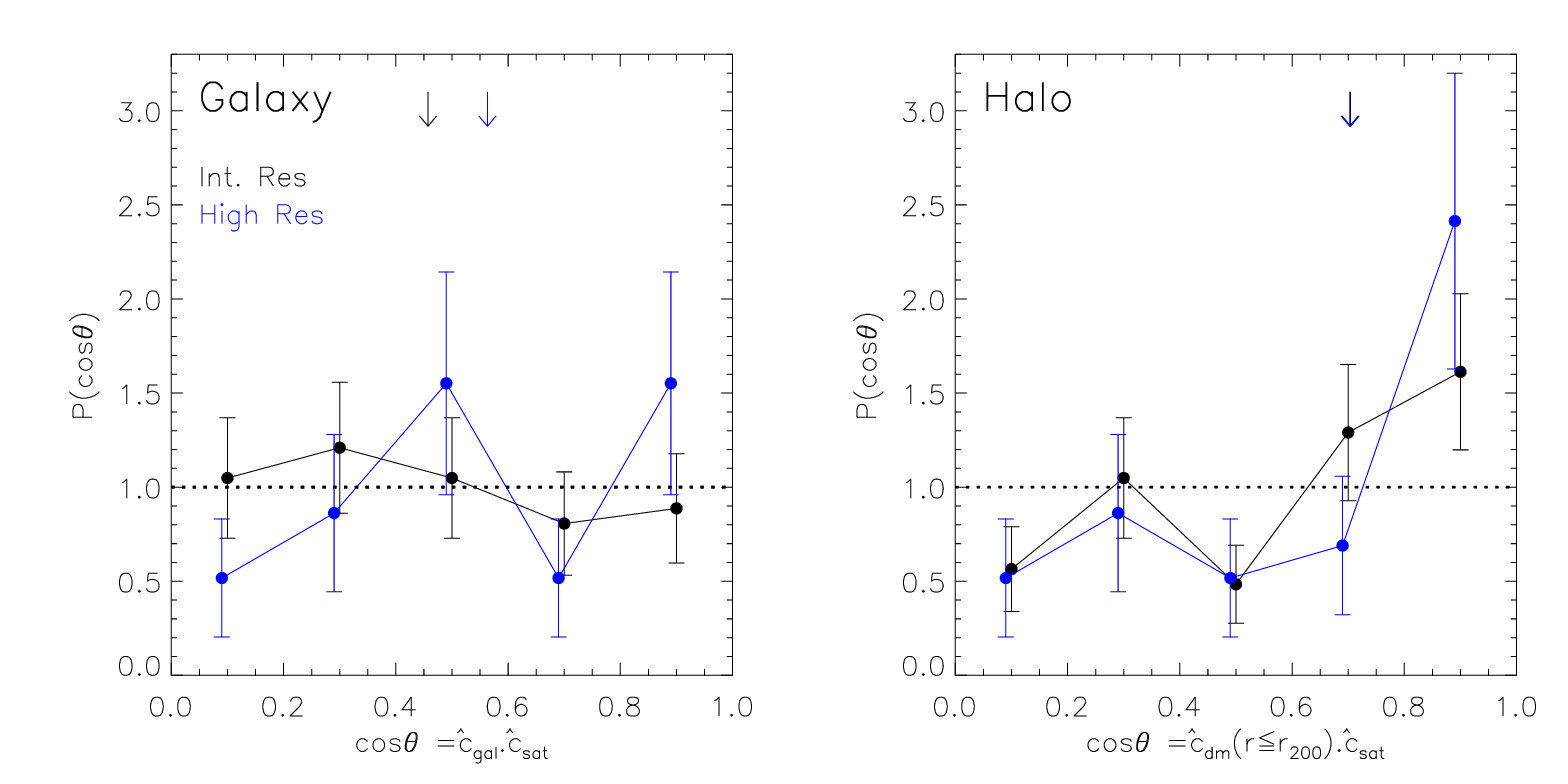}
  
   \caption{\small The orientation of the short axes of the satellite
    distribution relative to the short axes of the parent galaxy and dark matter halo (defined within $r_{200}$). The 10 brightest satellites
    within $r_{200}$ are used to compute the shapes of the satellite
    distribution. Downward pointing arrows denote the median of the
    distributions and the dotted lines indicate a uniform
    distribution. The error bars denote Poisson uncertainties. The
    black and blue lines show the distributions in the intermediate
    and high resolution simulations respectively.}
   \label{fig:app2}
\end{figure*}

Our sample of simulated galaxies is drawn from the five
intermediate-resolution \textsc{gimic} simulations, which have been
run to $z=0$. In addition, there is one higher resolution \textsc{gimic}
simulation (the -$2\sigma$ region) available to $z=0$. This simulation
has eight times better resolution and allows us to assess the numerical convergence of our results. We have checked that the main results in this paper are unchanged in
this higher resolution \textsc{gimic} simulation. Here, we give
examples for two of our main results. 

In Fig. \ref{fig:app1} we we show the distributions of misalignments between the galaxy
and dark matter halo short
axes both for $r\le 0.1r_{200}$ (left panel) and for
$r \le r_{200}$ (right panel) for the high resolution \text{gimic}
runs. This is shown in the bottom panels of Fig. \ref{fig:histogram} in the
main text. There is very good agreement between the high and
intermediate resolution simulations. Our finding in the main text that
there can be significant misalignments between the galaxy and the
outer dark
matter halo is therefore robust to an increase in resolution.

In Fig. \ref{fig:app2} we show the distribution of alignments between the short axes of the
satellite systems and the short axes of the dark matter halo (for $r \le r_{200}$) and
the galaxy of their parent haloes (see Fig. \ref{fig:sats_shape_dmgal}
in the main text). We only consider the 10 brightest
satellites within $r_{200}$. The black and blue lines show the
distributions in the intermediate and high resolution simulations
respectively. Note that only the $-2\sigma$ volume is available at
$z=0$ in the high resolution simulations so we do not achieve the same
statistics as the intermediate resolutions runs. Approximately $30$
parent haloes have 10 or more satellites within $r_{200}$ in the high
resolution sample (c.f. $\sim 80$ in the intermediate resolution
sample). For both the intermediate and high resolution runs the satellite distribution
preferentially aligns in a plane perpendicular to the short axis of
the dark matter distribution and the satellites show no
preferential alignment relative to the galaxy. Thus, our conclusions
from Fig. \ref{fig:sats_shape_dmgal} in the main text are robust to an
increased resolution.

\end{appendix}

\label{lastpage}
\end{document}